\documentclass{PoS}
\usepackage{axodraw}
\usepackage{cite}

\def\bq{\begin{eqnarray}}
\def\eq{\end{eqnarray}}
\def\l{\langle}
\def\r{\rangle}
\def\eps{\varepsilon}

\title{Automated calculations for multi-leg processes}

\ShortTitle{Multi-leg processes}

\author{
\speaker{Stefan Weinzierl}
\\
        Institut f{\"u}r Physik, Universit{\"a}t Mainz, D - 55099 Mainz, Germany\\
        E-mail: \email{stefanw@thep.physik.uni-mainz.de}}

\abstract{
The search for signals of new physics at the forthcoming LHC experiments
involves the analysis of final states characterised by a high number
of hadronic jets or identified particles.
Precise theoretical predictions for these processes require the computation
of scattering amplitudes with a large number of external particles
and beyond leading order in perturbation theory.
The complexity of a calculation grows with the number of internal loops as
well as with the number of external legs.
Automatisation of at least next-to-leading order calculations for 
LHC processes is therefore a timely task.
I will discuss various approaches.
}

\FullConference{XI International Workshop on Advanced Computing and Analysis Techniques in Physics Research\\
		 April 23-27 2007\\
		 Amsterdam, the Netherlands}

\begin{document}


\section{Introduction}

\begin{figure}
\begin{center}
\includegraphics[bb= 60 450 500 770,width=0.8\textwidth]{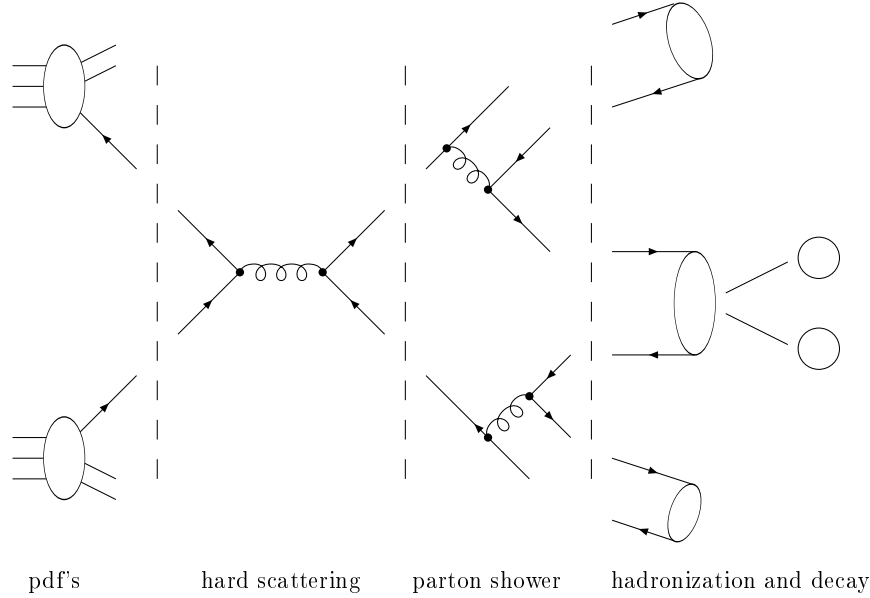}
\end{center}
\caption{A schematic and simplified description of an event in hadron-hadron collisions.}
\label{fig1}
\end{figure}

The Large Hadron Collider (LHC) is expected to start next year.
The physics program of the two main experiments ATLAS and CMS
aims at the discovery of the last missing particle predicted by the 
Standard Model -- the Higgs boson -- and at the search for signals of new physics beyond the Standard Model.
Common to these searches is the fact that the signal events have to be digged out from a bulk of background events.
The background events are due to Standard Model processes, mostly QCD processes which sometimes are accompanied
by additional electro-weak bosons.
Common to these searches is further the fact, that the final state is characterised 
by a high number of hadronic jets or 
identified particles.
A rough theoretical understanding of scattering processes at hadron colliders is sketched in 
fig.~\ref{fig1}. 
The initial protons do not take part as a single entity in the scattering process, 
instead the basic constituents of the proton -- partons like quarks and gluons --
enter the hard scattering process.
The probability of finding a specific parton inside the proton is described by
parton distribution functions (pdf's).
The parton distribution functions are non-perturbative objects and therefore at present cannot be calculated from theory.
However, they are universal and can be measured and extracted from one experiment and then used as input data 
for other experiments.
In particular, the HERA experiment at DESY contributed significantly to our knowledge of the
parton distribution functions.
As a technical detail, the parton distribution functions depend on a scale. In simplified terms, they are measured
at a scale $Q_0$, but used as input data at a different scale $Q_1$. 
In this context it is worth to note that the variation with the scale of the parton distribution functions
can be calculated within perturbation theory.

The hard scattering process can be formulated entirely in terms of the fundamental fields (quarks, gluons, ...) 
of the Standard Model.
It is calculable in perturbation theory.
Attached to the hard scattering process are parton showers, where the partons of the hard scattering event
radiate off additional collinear or soft partons.
From this additional radiation originates the observed hadronic jets, i.e. bunches of particles moving 
in the same direction.
The parton shower increases significantly the number of partons in the event.
In principle, the parton shower is governed to a large extend by perturbation theory.
But due to our limited computational abilities we are forced to replace the full matrix elements
by approximations. These approximations are based on the observation, that matrix elements are enhanced
in soft and collinear regions.

After the parton shower, the partons convert to hadrons. Like for the parton distribution functions
this is non-perturbative physics. In practise, models like the string model or the cluster model 
are used to mimic the formation of hadrons.
If unstable hadrons are formed, they subsequently decay.

This simplified picture of an event has to be completed with additional complications due to 
the underlying event, multiple interactions and pile-ups.
As the hard scattering breaks up the initial protons, the proton remnants are not colour-neutral.
The underlying event describes the interactions of the proton remnants.
Usually the underlying event will produce activity in the detector along the beam pipe.
However, it can happen that more than one pair of partons of the initial protons undergo a hard scattering.
This goes under the name of multiple interactions.
Finally, there is the possibility that more than one proton-proton scattering takes place in a single
bunch crossing. These are pile-up events.

As can be seen, the theoretical description of a single event is rather involved and 
requires in particular knowledge on non-perturbative physics.
As this knowledge is not available, we have to content ourselves to models and approximations.
This will limit the accuracy and precision of theoretical predictions.
Fortunately, for a specific class of observables we can do better:
This is the class of infrared-safe observables.
Infrared safeness implies that the value of an observable does not change, if additional soft or
collinear particles are added to an event:
\bq
 \lim\limits_{\mbox{\footnotesize $r$ partons unresolved}} O_{n+r}\left( p_1, ..., p_{n+r} \right)
 & = &
 O_n\left(p_1', ..., p_n' \right)
\eq
Here, $O_n\left(p_1, ..., p_n \right)$ denotes the value of the observable for an event with
particles with four-momenta $p_1$, ..., $p_n$.
Infrared-safe observables depend only mildly on showering and hadronisation and can be calculated
reliably in perturbation theory.
At hadron colliders we need of course the additional non-perturbative information on the 
parton distribution functions, but these quantities have been measured and are available.
The master formula for the calculation of an observable is given by
\bq
\l O \r & = &
             \sum\limits_{a,b}
             \int dx_1 \; f_a(x_1)
             \int dx_2 \; f_b(x_2)
             \;\;\;
             \frac{1}{2 K(s)}
             \;\;\;
             \frac{1}{\left( 2 J_1+1 \right)}
	       \frac{1}{\left( 2 J_2+1 \right)} \frac{1}{n_1 n_2}
 \nonumber \\
 & &
             \times
             \sum\limits_n
             \int d\phi_{n-2}
             O\left(p_1,...,p_n\right)
             \;\;\;
             \left| {\cal A}_n \right|^2.
\eq
The various ingredients of this formula are:
The parton distribution functions, i.e. the
probability of finding a parton $i$ with momentum fraction $x$ inside
the parent hadron $h$ are denoted by $f_i(x)$.
$2K(s)$ is the flux factor, equal to two times the center-of-mass energy squared
of the two incoming partons.
$1/(2J_1+1)/(2J_2+1)/n_1/n_2$ corresponds
to an averaging over the spins and colour degrees of freedom for the initial particles.
The second sum is over the number of final-state particles, the integral is over the phase space
of $(n-2)$ final-state particles.
The most important ingredient of this formula is the matrix element squared $| {\cal A}_n |^2$
for $n$ particles, $(n-2)$ in the final state, two in the initial state.
As can be seen from this formula, each event is weighted by the matrix element squared.

At high energies the strong coupling is small and the matrix element can be calculated in 
perturbation theory.
\begin{figure}
\begin{center}
\includegraphics[bb=130 530 475 655,width=0.8\textwidth]{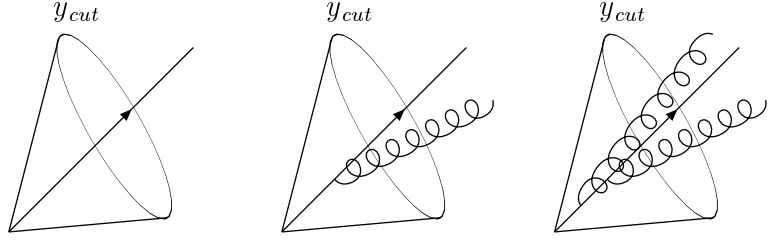}
\end{center}
\caption{Modelling of jets in perturbation theory. At leading-order a jet is modelled by a single parton,
at next-to-leading order either by one or two partons.
At next-to-next-to-leading order a jet is modelled by one, two or three partons.}
\label{fig2}
\end{figure}
For an observable, whose leading-order contribution in perturbation theory is given by
an $n$-parton amplitude, the following expansions are relevant for 
the calculation of the next-to-leading order (NLO) and
next-to-next-to-leading order (NNLO) predictions:
\bq
 & &
  \left| {\cal A}_n \right|^2
 =  
   \underbrace{
   \left| {\cal A}_n^{(0)} \right|^2
}_{\mbox{\footnotesize Born}} 
 + 
   \underbrace{
    2 \; \mbox{Re}
   \left(
             \left. {\cal A}_n^{(0)} \right.^\ast {\cal A}_n^{(1)} 
       \right)
}_{\mbox{\footnotesize one-loop}}
 + 
   \underbrace{
    2 \; \mbox{Re}
     \left(
             \left. {\cal A}_n^{(0)} \right.^\ast {\cal A}_n^{(2)} 
     \right)
           + \left|{\cal A}_n^{(1)} \right|^2 
}_{\mbox{\footnotesize two-loop and loop-loop}},
 \nonumber \\
 & &
  \left| {\cal A}_{n+1} \right|^2
 =  
   \underbrace{
\left| {\cal A}_{n+1}^{(0)} \right|^2
}_{\mbox{\footnotesize single emission}}
 + 
   \underbrace{
    2 \; \mbox{Re}
   \left(
          \left. {\cal A}_{n+1}^{(0)} \right.^\ast {\cal A}_{n+1}^{(1)} 
 \right)
}_{\mbox{\footnotesize loop+single emission}},
 \nonumber \\ 
 & &
  \left| {\cal A}_{n+2} \right|^2
 = 
   \underbrace{
   \left| {\cal A}_{n+2}^{(0)} \right|^2
}_{\mbox{\footnotesize double emission}}.
\eq
In this formulae
${\cal A}_n^{(l)}$ denotes an  amplitude with $n$ external particles and $l$ loops.
At leading-order only the Born amplitude ${\cal A}_n^{(0)}$ contributes.
At next-to-leading order we have in addition the contributions
from the one-loop amplitude ${\cal A}_n^{(1)}$ and the single emission contribution,
which is given by the Born amplitudes ${\cal A}_{n+1}^{(0)}$ with one additional parton.
These two contribution are of the same order with respect to the power counting of the
coupling.
As far as the phase space for the final-state particles is concerned,
these two contributions live on different phase spaces of different dimensions,
since the single emission contribution has one additional particle in the final state.
At next-to-next-to-leading order we have to include in addition the contributions
from the two-loop amplitude and the one-loop amplitude squared,
the one-loop amplitudes with a single additional emission and the Born amplitudes
with two additional emissions.
Up to next-to-next-to-leading order we include the radiation of up to two additional particles.
Therefore jets are modelled by one, two or three partons.
This is shown in fig.~\ref{fig2}.

\begin{figure}
\begin{center}
\includegraphics[width=7cm]{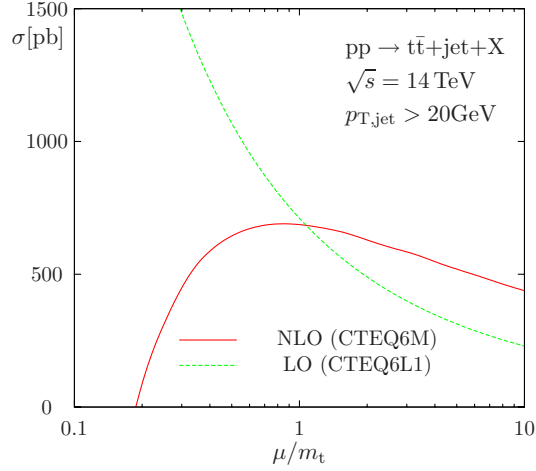}
\end{center}
\caption{Scale-dependence of the cross-section for the process $pp\rightarrow t\bar{t}+\mbox{jet}$.
The leading-order prediction shows a strong scale-dependence.
The next-to-leading order prediction reduces significantly the scale-dependence.
The plot is taken from \cite{Dittmaier:2007wz}.
}
\label{fig3}
\end{figure}

Why are the computations of higher-order corrections necessary ?
The answer is that we want to achieve a certain precision.
This is illustrated by a simple example.
Let us consider a pure QCD process with three hard partons in the final state.
The leading-order prediction is proportional to $\alpha_s^3$. It is a well-known fact
that the numerical value of the strong coupling depends on an arbitrary scale $\mu_{ren}$,
at leading order the formula reads
\bq
 \alpha_s & = & \frac{4\pi}{\beta_0 \ln \frac{\mu_{ren}^2}{\Lambda^2}},
 \;\;\;\;\;
 \beta_0 = 11 - \frac{2}{3} N_f,
 \;\;\;
 \Lambda \approx 165 \; \mbox{MeV}
 \;\;\;\mbox{for}\; N_f=5 \;
 \mbox{light flavours}.
\eq
As the choice of the scale is arbitrary, and since $\alpha_s$ enters the leading-order prediction
to the third power, this introduces a strong scale-dependence of the theoretical prediction
and therefore a large uncertainty.
The inclusion of higher-order corrections reduces this uncertainty.
This is shown for the example of $t\bar{t}+\mbox{jet}$ production at the LHC in fig.~\ref{fig3}.
This is an example where the leading-order prediction is proportional to 
$\alpha_s^3$.
Shown in fig.~\ref{fig3} is the variation of the theoretical prediction with the scale $\mu$.
Apart from the scale entering the formula for the strong coupling, which is called the
renormalisation scale there is in addition also a scale, at which the parton distribution functions
are evaluated. The latter is called the factorisation scale.
In the plot both scale have been identified,
\bq
 \mu = \mu_{ren} = \mu_{fact}.
\eq
As can be seen, the inclusion of the next-to-leading order correction reduces
significantly the scale-dependence
and clearly motivates the quest for the computation of higher-order corrections.

What are the objectives for LHC physics ?
Let us now look more closely where precision calculations are needed for LHC physics.
There are three important points to mention:
\begin{itemize}
\item In order to predict absolute rates to a good precision, we have to know
the input parameters of a theoretical calculation to a high precision.
The most important input parameters are the strong coupling $\alpha_s$
and the parton distribution functions.
These parameters can be extracted from other experiments, for example 
the strong coupling can be extracted from the process
$e^+ e^- \rightarrow \mbox{3 jets}$ at LEP, the pdf's are
measured at HERA.
The experimental precision has to be matched with the corresponding precision of the
theoretical calculation, requiring the inclusion of the NNLO corrections.
This in turn requires a calculation for $e^+ e^- \rightarrow \mbox{3 jets}$ at NNLO
and the computation of the three-loop splitting functions for the evolution of the
pdf's.
\item Standard candles at the LHC: 
For a few standard hard $pp$ processes like the production of
$W$, $Z$, top, Higgs or jets we would like to have a precise prediction from theory,
again requiring a NNLO computation.
Higgs production is of clear interest for the search of the last undiscovered
particle predicted by the Standard Model.
The other standard processes are useful to measure fundamental quantities like
the $W$- or top-mass to a better precision or can be used by the experimentalists
to understand their detector.
\item Finally one would like to have reliable 
predictions for multi-particle final states that occur at high rates and form 
background to new physics. This implies next-to-leading order calculations.
\begin{table}
\begin{center}
\begin{tabular}{lll}
& process & relevant for \\
\hline
 & & \\
1. & $p p \rightarrow V V + \mbox{jet}$ & $t \bar{t} H$, new physics \\
2. & $p p \rightarrow H + \mbox{2 jets}$ & Higgs production by vector boson fusion \\
3. & $p p \rightarrow t \bar{t} b \bar{b}$ & $t \bar{t} H$\\
4. & $p p \rightarrow t \bar{t} + \mbox{2 jets}$ & $t \bar{t} H$ \\
5. & $p p \rightarrow V V b \bar{b}$ & $VBF \rightarrow H \rightarrow VV$, $t \bar{t} H$, new physics \\
6. & $p p \rightarrow V V + \mbox{2 jets}$ & $VBF \rightarrow H \rightarrow VV$ \\
7. & $p p \rightarrow V + \mbox{3 jets}$ & various new physics signatures \\
8. & $p p \rightarrow V V V$ & SUSY\\
\end{tabular}
\end{center}
\caption{The experimenter's wish list at Les Houches 2005 \cite{Buttar:2006zd}. $V$ denotes either one of the
electro-weak bosons: photon, $Z$-boson or $W$-boson.
}
\label{table1}
\end{table}
Examples for such processes have been collected in the 
experimenter's wish list at Les Houches workshop in 2005 and are shown in table~\ref{table1}.
\end{itemize}

What is the state-of-the-art for multi-leg processes ?
It is clear that the complexity of a calculation increases with the order in perturbation
theory as well as with the number of external legs.
Therefore we expect that if we go up in the order of perturbation theory the available calculations are restricted
to fewer final state particles. 
Let us first consider leading order calculations.
At this order there are techniques for the automated calculation of the (leading order)
matrix elements 
\cite{Berends:1987me,Berends:1989ie,Berends:1990ax,Caravaglios:1995cd,Caravaglios:1998yr,Draggiotis:1998gr,Draggiotis:2002hm,Papadopoulos:2005ky}
as well as techniques for the efficient integration over phase space \cite{Kleiss:1986gy,vanHameren:2002tc}.
Several computer programs like Madgraph/Madevent \cite{Maltoni:2002qb,Stelzer:1994ta,Murayama:1992gi}, Sherpa/Amegic++ \cite{Krauss:2001iv}, 
Helac/Phegas \cite{Kanaki:2000ey,Papadopoulos:2000tt},
Comphep \cite{Pukhov:1999gg}, Grace \cite{Yuasa:1999rg} or Alpgen \cite{Mangano:2002ea}
are available, which implement these techniques and can be used
to obtain leading order predictions for processes with a rather high number of final
state particles.

If we now consider the next order in perturbation theory, we observe that
there many NLO calculation for $2 \rightarrow 2$ processes at
hadron colliders, but only a few for $2 \rightarrow 3$ processes.
Fully differential numerical programs for $2 \rightarrow 3$ processes exist 
for example for
$p p \rightarrow \mbox{3 jets}$ \cite{Kilgore:1996sq,Nagy:2001fj,Nagy:2003tz},
$p p \rightarrow V + \mbox{2 jets}$ \cite{Campbell:2000bg,Campbell:2002tg,Campbell:2003hd,Campbell:2005zv,Campbell:2006cu},
$p p \rightarrow t \bar{t} H$ \cite{Beenakker:2002nc,Dawson:2003zu},
$p p \rightarrow H + \mbox{2 jets}$ \cite{DelDuca:2001eu,DelDuca:2001fn,Campbell:2006xx},
$p p \rightarrow t \bar{t} + \mbox{jet}$ \cite{Dittmaier:2007wz}
and 
$p p \rightarrow Z Z Z$ \cite{Lazopoulos:2007ix}.
Of comparable complexity is the NLO calculation for the production of two vector bosons in 
vector-boson-fusion \cite{Jager:2006zc,Jager:2006cp,Bozzi:2007ur}.
On the other side there aren't at present any NLO programs for the LHC
with more than 3 hard particles in the final state.
NLO programs with four particles in the final state are available for electron-positron annihilation for
$e^+ e^- \rightarrow \mbox{4 fermions}$ \cite{Denner:2005es,Denner:2005fg}
and $e^+ e^- \rightarrow \mbox{4 jets}$
\cite{Dixon:1997th,Nagy:1997yn,Campbell:1998nn,Weinzierl:1999yf}.
As can be seen, for NLO computations our capabilities are already much more restricted with respect
to the number of final state particles.
For processes with only two or three final state particles tools which help with the automatisation
of the computation like the combination FeynArts, FormCalc, Looptools 
\cite{Kublbeck:1990xc,Hahn:2000kx,Hahn:1998yk,vanOldenborgh:1990yc}
or the Grace package \cite{Belanger:2003sd}
are available.

Finally, we consider NNLO calculations. Here, fully differential predictions for the LHC are available only for a few
selected $2 \rightarrow 1$ and $2 \rightarrow 2$ processes, like Drell-Yan \cite{Anastasiou:2003yy}, $W$-production \cite{Anastasiou:2003ds}
or Higgs production \cite{Anastasiou:2005qj,Anastasiou:2004xq,Anastasiou:2002yz,Harlander:2002wh,Ravindran:2003um,Harlander:2001is,Catani:2001ic}.
For electron-positron annihilation NNLO predictions are available for
$e^+ e^- \rightarrow \mbox{2 jets}$ \cite{Anastasiou:2004qd,Weinzierl:2006ij,Weinzierl:2006yt}
and the thrust distribution \cite{Ridder:2007bj}.

What are the bottle-necks ? As can be seen from this summary of the state-of-the-art, higher-order corrections are 
limited to processes with not too many particles in the final state.
Let us now look into the difficulties, which prohibit a straight-forward automatisation of the computation of higher
order corrections for processes with many particles in the final state.
These difficulties can be grouped into three categories:
\begin{itemize}
\item Length: Perturbative calculations lead to expressions with a huge number of terms.
\item Integrals: At one-loop and beyond, the occurring integrals cannot be simply looked up in an integral table.
\item Divergences: At NLO and beyond, infrared divergences occur in intermediate stages, if massless particles are involved.
\end{itemize}
The first complication -- lengthy expressions -- affects already leading-order computations.
I will discuss methods to handle this problem in section \ref{sect:length}.
The second and the third item occur for the first time in NLO computations.
They are discussed in section \ref{sect:loop} and section \ref{sect:IR}, respectively.


\section{Managing lengthy expressions}
\label{sect:length}

It is a well-known fact that the complexity of a calculation based on Feynman diagrams
growth factorially with the number of external particles.
As an example consider the all-gluon amplitude at tree level.
\begin{table}
\begin{center}
\begin{tabular}{l|r|r|r|r|r|r|r}
 n & 2 & 3 & 4 & 5 & 6 & 7 & 8 \\
\hline
 diagrams & 4 & 25 & 220 & 2485 & 34300 & 559405 & 10525900 \\
\end{tabular}
\end{center}
\caption{Number of Feynman diagrams contributing to $g g \rightarrow n g$ at tree level.}
\label{table2}
\end{table}
Table \ref{table2} shows the number of diagrams contributing to the process $g g \rightarrow n g$ at tree level.
All diagrams involve three- and four-gluon vertices.
The Feynman rules for these vertices blow-up the corresponding expressions even further:
\bq
\begin{picture}(100,35)(0,55)
\Vertex(50,50){2}
\Gluon(50,50)(50,80){3}{4}
\Gluon(50,50)(76,35){3}{4}
\Gluon(50,50)(24,35){3}{4}
\end{picture}
 & = &
g f^{abc} \left[ (k_{3} - k_{2})_{\mu} g_{\nu \lambda} 
               + (k_{1} - k_{3})_{\nu} g_{\lambda \mu} 
               + (k_{2} - k_{1})_{\lambda} g_{\mu \nu} \right],
 \nonumber \\
\begin{picture}(100,75)(0,50)
\Vertex(50,50){2}
\Gluon(50,50)(71,71){3}{4}
\Gluon(50,50)(71,29){3}{4}
\Gluon(50,50)(29,29){3}{4}
\Gluon(50,50)(29,71){3}{4}
\end{picture}
 & = &
- i g^{2} \left[ f^{abe} f^{ecd} \left( g_{\mu \lambda} g_{\nu \rho} - g_{\mu \rho} g_{\nu \lambda} 
                                 \right) 
               + f^{ace} f^{ebd} \left( g_{\mu \nu} g_{\lambda \rho} - g_{\mu \rho} g_{\lambda \nu} \right) 
 \right. \nonumber \\
 & & \left.
         + f^{ade} f^{ecb} \left( g_{\mu \nu} g_{\lambda \rho} - g_{\mu \lambda} g_{\nu \rho} \right)
\right].
\nonumber \\
\eq
For the computation of observables we have to square the amplitude and to sum over all
spins or helicities.
For gluons we replace the polarisation sum by
\bq
\sum\limits_{\lambda} \eps_\mu^\ast(k,\lambda) \eps_\nu(k,\lambda)
 & = & 
 - g_{\mu\nu} + \frac{k_\mu n_\nu + n_\mu k_\nu}{k n} - n^2 \frac{k_\mu k_\nu}{(k n )^2}.
\eq 
In this formula, $n_\mu$ is an arbitrary four-vector. 
If the amplitude consists of $O(N)$ terms, squaring the amplitude will produce $O(N^2)$ terms.
From this example it is clear, that the number of terms in intermediate expressions of
a calculation based on Feynman diagrams growth dramatically with the number of
external legs. 
As already mentioned, this problem occurs already at tree level.


\subsection{Computer algebra}
\label{sect_computer_algebra}

As calculations for multi-particle final states tend to involve lengthy intermediate expressions,
computer algebra has become an essential tool.
In fact, particle physics is and has been a driving force for the development of computer algebra systems.
Quite a few computer algebra systems have their roots within the high energy
physics community or strong links with them:
REDUCE,
SCHOONSHIP,
MATHEMATICA,
FORM \cite{Vermaseren:2000nd}
or GiNaC \cite{Bauer:2000cp,Vollinga:2005pk}, 
to name only a few.
In most cases the requirements on a computer-algebra system for
computer-intensive symbolic calculations
in particle physics can be summarised by:
\begin{itemize}
\item The computer algebra system has to provide 
basic operations like addition, multiplication, sorting, etc..
\item Specialised code for the solution of a particular problem is 
usually written by the user. The computer algebra system has to provide
a convenient programming language.
\item There is actually no need for a system which knows ``more'' than the user.
\end{itemize}
The most widely used computer algebra systems in the community are 
MATHEMATICA, MAPLE, REDUCE, FORM and GiNaC.
The first three are commercial programs, FORM and GiNAC are non-commercial and 
freely available\footnote{FORM is available at http://www.nikhef.nl/\~{}form, GiNaC is available at
http://www.ginac.de.}.
Below there are two small example
programs in FORM and GiNaC for the calculation of
\bq 
 \mbox{Tr} \; p\!\!\!/_1 p\!\!\!/_2 p\!\!\!/_3 p\!\!\!/_4
 & = &
 4 \left( p_1 \cdot p_2 \right) \left( p_3 \cdot p_4 \right)
 - 4 \left( p_1 \cdot p_3 \right) \left( p_2 \cdot p_4 \right)
 + 4 \left( p_1 \cdot p_4 \right) \left( p_2 \cdot p_3 \right).
\eq
The example in FORM reads:
{\footnotesize
\begin{verbatim}
* Example program for FORM

V p1,p2,p3,p4;

L  res = g_(1,p1)*g_(1,p2)*g_(1,p3)*g_(1,p4);

trace4,1;

print;

.end
\end{verbatim}
}
GiNaC is a C++ library, which provides capabilities for symbolic calculations within the C++
programming language.
The corresponding example in GiNaC reads:
{\footnotesize
\begin{verbatim}
#include <iostream>
#include <ginac/ginac.h>

using namespace std;
using namespace GiNaC;

int main()
{
  varidx mu(symbol("mu"),4), nu(symbol("nu"),4), 
         rho(symbol("rho"),4), sigma(symbol("sigma"),4);

  symbol p1("p1"), p2("p2"), p3("p3"), p4("p4");

  ex res = dirac_gamma(mu,1)*dirac_gamma(nu,1)
    *dirac_gamma(rho,1)*dirac_gamma(sigma,1)
    *indexed(p1,mu.toggle_variance())*indexed(p2,nu.toggle_variance())
    *indexed(p3,rho.toggle_variance())*indexed(p4,sigma.toggle_variance());

  res = dirac_trace(res,1);

  res = res.expand();
  res = res.simplify_indexed();

  cout << res << endl;
}
\end{verbatim}
}
Computer algebra is an essential tool, but a brute force application alone will still produce
lengthy expressions, which are slow and potentially unstable, when evaluated numerically.


\subsection{Quantum number management}
\label{sect:quantum_numbers}

In order to keep the size of intermediate expressions under control, a divide-and-conquer strategy
has been proven useful: One divides the quantity to be calculated into smaller pieces and calculates the
small pieces separately.
This approach is also called ``quantum number management''.
One first observes that it is not necessary to square the amplitude and sum over the spins and helicities
analytically. It is sufficient to do this numerically. This avoids obtaining $O(N^2)$ terms from
an expression with $O(N)$ terms.
The individual amplitudes have to be calculated in a helicity or spin basis. This is discussed in 
section \ref{sect:helicity}.
The second observation is related to the fact, that individual helicity amplitudes can be decomposed
into smaller gauge-invariant pieces, called partial amplitudes. 
This is discussed in section \ref{sect:colour}.
These partial amplitudes can be calculated for a given helicity configuration without reference
to Feynman diagrams, as discussed in section \ref{sect:Berends_Giele}.
Finally, for specific helicity configurations, compact formul{\ae} are known, which are summarised in 
section \ref{sect:parke_taylor}.
Supersymmetric identities provide relations between amplitudes with different particle contents,
therefore only some of them need to be calculated. This is discussed in section \ref{sect:susy}.

\subsubsection{Helicity amplitudes}
\label{sect:helicity}

The computation of helicity amplitudes requires the choice of a helicity (or spin) basis.
This is straightforward for massless fermions. The two-component Weyl spinors
provide a convenient basis:
\bq
 | p \pm \rangle & = & \frac{1}{2} \left( 1 \pm \gamma_5 \right) u(p).
\eq
In the literature there are different notations for Weyl spinors.
Apart from the bra-ket-notation there is the notation with dotted and un-dotted indices:
The relation between the two notations is the following:
\bq
|p+\rangle = p_B,          & & \langle p+| = p_{\dot{A}}, \nonumber \\
|p-\rangle = p^{\dot{B}},  & & \langle p-| = p^A. 
\eq
Spinor products are defined as
\bq
\langle p q \rangle = \langle p - | q + \rangle, 
 & &
\left[ p q \right] = \langle p + | q - \rangle,
\eq
and take value in the complex numbers. 
It was a major break-through, when it was realised that also gluon polarisation vectors
can be expressed in terms of 
two-component Weyl spinors\cite{Berends:1981rb,DeCausmaecker:1982bg,Gunion:1985vc,Kleiss:1986qc,Xu:1987xb,Gastmans:1990xh}.
The polarisation vectors of external gluons can be chosen as
\bq
\label{gluon_pol_onshell}
\eps_{\mu}^{+}(k,q) = \frac{\langle q-|\gamma_{\mu}|k-\rangle}{\sqrt{2} \langle q- | k + \rangle},
 & &
\eps_{\mu}^{-}(k,q) = \frac{\langle q+|\gamma_{\mu}|k+\rangle}{\sqrt{2} \langle k + | q - \rangle},
\eq
where $k$ is the momentum of the gluon and $q$ is an arbitrary light-like reference momentum.
The dependence on the arbitrary reference momentum $q$ will drop out in gauge
invariant quantities. 
The polarisation sum is that of an light-like axial gauge:
\bq
\sum\limits_{\lambda = \pm} \varepsilon_\mu^\lambda(k,q) \left( \varepsilon_\nu^\lambda(k,q) \right)^\ast
& = & - g_{\mu\nu} + \frac{k_\mu q_\nu + k_\nu q_\mu}{k \cdot q}.
\eq
Changing the reference momentum will give a term proportional to the
momentum of the gluon:
\bq
\varepsilon^+_\mu(k,q_1) -\varepsilon^+_\mu(k,q_2)& = & \sqrt{2} \frac{\langle q_1 q_2 \rangle}
{\langle q_1 k \rangle \langle k q_2 \rangle} k_\mu .
\eq
For massive fermions we can take the spinors as \cite{Schwinn:2005pi,Rodrigo:2005eu}
\bq
\label{eq:spinors} 
 u(\pm) = \frac{1}{\l p^\flat \mp | q \pm \r} \left( p\!\!\!/ + m \right) | q \pm \r,
 & &
\bar{u}(\pm) = \frac{1}{\l q \mp | p^\flat \pm \r} \l q \mp | \left( p\!\!\!/ + m \right),
 \nonumber \\
 v(\pm) = \frac{1}{\l p^\flat \mp | q \pm \r} \left( p\!\!\!/ - m \right) | q \pm \r,
&&
\bar{v}(\pm) = \frac{1}{\l q \mp | p^\flat \pm \r} \l q \mp | \left( p\!\!\!/ - m \right).
\eq
Here $p^\flat$ is a light-like four vector obtained through
\bq
\label{projection_null}
 p^\flat & = & p - \frac{p^2}{2 p \cdot q} q.
\eq
$q$ denotes again an arbitrary light-like reference momentum and is related
to the quantisation axis of the spin for 
the massive fermion.
In contrast to the gluon case 
individual amplitudes with label $+$ or $-$ will depend on the choice of the reference momentum $q$.

\subsubsection{Colour decomposition}
\label{sect:colour}

Throughout this article I use the normalisation
\bq
 \mbox{Tr}\;T^a T^b & = & \frac{1}{2} \delta^{a b}
\eq
for the colour matrices.
Amplitudes in QCD may be decomposed into group-theoretical factors (carrying the colour structures)
multiplied by kinematic functions called partial amplitudes
\cite{Cvitanovic:1980bu,Berends:1987cv,Mangano:1987xk,Kosower:1987ic,Bern:1990ux}. 
These partial amplitudes do not contain any colour information and are gauge-invariant objects. 
In the pure gluonic case tree level amplitudes with $n$ external gluons may be written in the form
\bq
{\cal A}_{n}(1,2,...,n) & = & g^{n-2} \sum\limits_{\sigma \in S_{n}/Z_{n}} 2 \; \mbox{Tr} \left(
 T^{a_{\sigma(1)}} ... T^{a_{\sigma(n)}} \right)
 A_{n}\left( \sigma(1), ..., \sigma(n) \right), 
\eq
where the sum is over all non-cyclic permutations of the external gluon legs.
The quantities $A_n(\sigma(1),...,\sigma(n))$, called the partial amplitudes, contain the kinematic information.
They are colour-ordered, e.g. only diagrams with a particular cyclic ordering of the gluons contribute.
The colour decomposition is obtained by replacing the structure constants $f^{abc}$
by
\bq
 i f^{abc} & = & 2 \left[ \mbox{Tr}\left(T^a T^b T^c\right) - \mbox{Tr}\left(T^b T^a T^c\right) \right] 
\eq
which follows from $ \left[ T^a, T^b \right] = i f^{abc} T^c$.
The resulting traces and strings of colour matrices can be further simplified with
the help of the Fierz identity :
\bq
 T^a_{ij} T^a_{kl} & = &  \frac{1}{2} \left( \delta_{il} \delta_{jk}
                         - \frac{1}{N} \delta_{ij} \delta_{kl} \right).
\eq
The colour decomposition for a tree amplitude with a pair of quarks is
\bq
{\cal A}_{n+2}(q,1,2,...,n,\bar{q}) & = & g^{n} \sum\limits_{S_n} \left( T^{a_{\sigma(1)}} ... T^{a_{\sigma(n)}} \right)_{i_q j_{\bar{q}}}
A_{n+2}(q,\sigma(1),\sigma(2),...,\sigma(n),\bar{q}). 
\eq
where the sum is over all permutations of the gluon legs. 
In general, the colour factors are combinations of open
\bq
 \left( T^{a_1} ... T^{a_n} \right)_{i_q j_{\bar{q}}}
\eq
and closed strings
\bq
\mbox{Tr} \left( T^{b_1} ... T^{b_m} \right)
\eq
of colour matrices. These building blocks form a basis in colour space.
The choice of the basis for the colour structures is not unique, and several proposals
for bases can be found in the literature \cite{DelDuca:1999rs,Maltoni:2002mq,Weinzierl:2005dd}.
A second useful basis is the colour-flow basis:
This basis is obtained by replacing every contraction over an index in the adjoint
representation by two contractions over
indices $i$ and $j$ in the fundamental representation:
\bq
V^a E^a & = & V^a \delta^{ab} E^b = V^a \left( 2 T^a_{ij} T^b_{ji} \right) E^b
 =  \left( \sqrt{2} T^a_{ij} V^a \right) \left( \sqrt{2} T^b_{ji} E^b \right).
\eq
In this representation the colour decomposition of the pure gluon amplitude is given by
\bq
{\cal A}_{n}(1,2,...,n) & = & \left(\frac{g}{\sqrt{2}}\right)^{n-2} \sum\limits_{\sigma \in S_{n}/Z_{n}} 
 \delta_{i_{\sigma_1} j_{\sigma_2}} \delta_{i_{\sigma_2} j_{\sigma_3}} 
 ... \delta_{i_{\sigma_n} j_{\sigma_1}}  
 A_{n}\left( \sigma_1, ..., \sigma_n \right).
\eq

\subsubsection{Off-shell recurrence relations}
\label{sect:Berends_Giele}

Berends-Giele type recurrence relations \cite{Berends:1987me,Kosower:1989xy}
build partial amplitudes from smaller building blocks, usually
called colour-ordered off-shell currents.
Off-shell currents are objects with $n$ on-shell legs and one additional leg off-shell.
Momentum conservation is satisfied. It should be noted that
off-shell currents are not gauge-invariant objects.
Recurrence relations relate off-shell currents with $n$ legs 
to off-shell currents with fewer legs.
The recursion starts with $n=1$:
\bq
J^\mu(k_1) & = & \eps^\mu(k_1,q).
\eq
$\eps^\mu$ is the polarisation vector of the gluon and $q$ an arbitrary light-like reference momentum.
The recursive relation states that in the pure-gluon off-shell current
a gluon couples to other gluons only via the three- or four-gluon
vertices :
\bq
\label{Berends_Giele_recursion}
 J^\mu(k_1^{\lambda_1},...,k_n^{\lambda_n}) & = & 
 \frac{-i}{K^2_{1,n}} 
 \left[ 
        \sum\limits_{j=1}^{n-1} V_3^{\mu\nu\rho}(-K_{1,n},K_{1,j},K_{j+1,n})
                                J_\nu(k_1^{\lambda_1},...,k_j^{\lambda_j}) J_\rho(k_{j+1}^{\lambda_{j+1}},...,k_n^{\lambda_n}) 
 \right. \nonumber \\
 & & \left. 
        + \sum\limits_{j=1}^{n-2} \sum\limits_{l=j+1}^{n-1} V_4^{\mu\nu\rho\sigma} 
            J_\nu(k_1^{\lambda_1},...,k_j^{\lambda_j}) J_\rho(k_{j+1}^{\lambda_{j+1}},...,k_l^{\lambda_l}) J_\sigma(k_{l+1}^{\lambda_{l+1}},...,k_n^{\lambda_n}) 
     \right],
\eq
where
\bq
K_{i,j} & = & k_i + k_{i+1} + ... + k_j 
\eq
\begin{figure}
\begin{center}
\includegraphics[bb=115 525 590 660,width=0.8\textwidth]{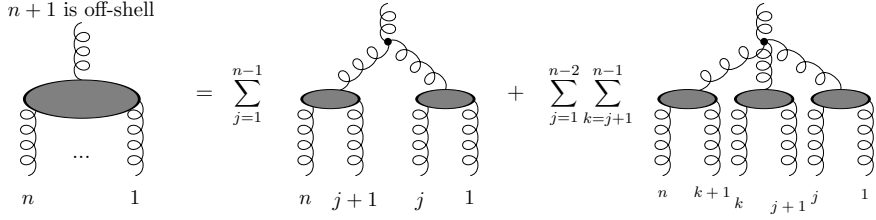}
\end{center}
\caption{Off-shell recurrence relation: In an off-shell current particle $n+1$ is kept off-shell.
This allows to express an off-shell current with $n$ on-shell legs in terms of currents with fewer legs.}
\label{fig4}
\end{figure}
and $V_3$ and $V_4$ are the colour-ordered three-gluon and four-gluon vertices
\bq
\label{Feynman_rules}
 V_3^{\mu\nu\rho}(k_1,k_2,k_3) 
 & = & 
 i \left[
          g^{\mu\nu} \left( k_1^\rho - k_2^\rho \right)
        + g^{\nu\rho} \left( k_2^\mu - k_3^\mu \right)
        + g^{\rho\mu} \left( k_3^\nu - k_1^\nu \right)
   \right],
 \nonumber \\
 V_4^{\mu\nu\rho\sigma} & = & i \left( 2 g^{\mu\rho} g^{\nu\sigma} - g^{\mu\nu} g^{\rho\sigma} -g^{\mu\sigma} g^{\nu\rho} \right).
\eq
The recurrence relation is shown pictorially in fig.~\ref{fig4}.
The gluon current $J_\mu$ is conserved:
\bq
\left( \sum\limits_{i=1}^n k_i^\mu \right) J_\mu & = & 0.
\eq
From an off-shell current one easily recovers the on-shell amplitude by 
removing the extra propagator,
taking the leg $(n+1)$ on-shell
and contracting with the appropriate polarisation vector.

\subsubsection{Parke-Taylor formul\ae}
\label{sect:parke_taylor}

The partial amplitudes have 
for specific helicity combinations remarkably
simple analytic formula or vanish altogether.
For the all-gluon tree amplitude one finds
\bq
A_{n}(1^+,2^+,...,n^+) & = & 0,
 \nonumber \\
A_{n}(1^+,2^+,...,j^-,...,n^+) & = & 0,
 \nonumber \\
A_{n}(1^+,2^+,...,j^-,...,k^-,...,n^+) 
 & = & i \left( \sqrt{2} \right)^{n-2} 
 \frac{\langle j k \rangle^4}{\langle 1 2 \rangle ... \langle n 1 \rangle}.
\eq
The partial amplitudes where all gluons have positive helicities, or where all gluons except one
have positive helicities vanish.
The first non-vanishing result is obtained for 
the $n$-gluon amplitude with $n-2$ gluons of positive helicity and $2$ gluons of negative helicity.
It is given by a remarkable simple formula. Note that this formula holds for all $n$.
An amplitude with $n-2$ gluons of positive helicity and $2$ gluons of negative helicity
is called a
maximal-helicity violating amplitude (MHV amplitude).
Obviously, we find similar formul{\ae} if we exchange all positive and negative helicities:
\bq
A_{n}(1^-,2^-,...,n^-) & = & 0,
 \nonumber \\
A_{n}(1^-,2^-,...,j^+,...,n^-) & = & 0,
 \nonumber \\
A_{n}(1^-,2^-,...,j^+,...,k^+,...,n^-) 
 & = & i \left( \sqrt{2} \right)^{n-2} 
 \frac{[ k j ]^4}{[1 n ] [n (n-1)] ... [ 2 1 ]}.
\eq
These formul{\ae} have been conjectured by Parke and Taylor \cite{Parke:1986gb}
and have been proven by
Berends and Giele \cite{Berends:1987me}.

\subsubsection{Supersymmetric relations}
\label{sect:susy}

After removing the colour factors, QCD at tree-level may be viewed as an 
effective supersymmetric theory \cite{Grisaru:1976vm,Grisaru:1977px,Parke:1985pn,Reuter:2002gn,Schwinn:2006ca},
where the quarks and the gluons form a super-multiplet (a $N=1$ vector super-multiple).
Let us denote by
\bq
Q_{\mbox{\tiny SUSY}} & = & \theta \left( q^A Q_A + \bar{q}_{\dot A}\bar Q^{\dot A} \right)
\eq
the SUSY generators contracted with two-component Weyl spinors $q^A$ and $\bar{q}_{\dot{A}}$ and
multiplied by a Grassmann number $\theta.$
In an unbroken supersymmetric theory, the supercharge 
annihilates the vacuum, and therefore
\bq
\left\langle 0 \left| \left[ Q_{\mbox{\tiny SUSY}}, \Phi_1 \Phi_2 ... \Phi_n \right] \right| 0 \right\rangle =
\sum\limits_{i=1}^n \left\langle 0 \left| \Phi_1 ... \left[ Q_{\mbox{\tiny SUSY}}, \Phi_i \right] ... \Phi_n \right| 0 \right\rangle =
0 
\eq
where the field $\Phi_i$ denotes either a gauge boson $g$ or a fermion $\Lambda$. 
The commutators are given by
\bq
\left[ Q_{\mbox{\tiny SUSY}}, g^\pm(k) \right] = \Gamma^\pm(k,q) \Lambda^\pm(k), & &
\left[ Q_{\mbox{\tiny SUSY}}, \Lambda^\pm(k) \right] = \Gamma^\mp(k,q) g^\pm(k),
\eq
with
\bq
 \Gamma^\pm(k,q) = \theta \langle q\pm | k\mp \rangle.
\eq
Let us now consider
\bq
 0 & = & 
 \left\langle 0 \left| \left[ Q, 
                              \Lambda_1^+ g_2^+ ... g_j^- ... g_{n-1}^+ g_n^- 
               \right] \right| 0 \right\rangle 
 \nonumber \\
 & = & \Gamma^-(p_1,q) A_n(g_1^+,g_2^+,...,g_j^-,...,g_{n-1}^+,g_n^- ) 
       - \Gamma^-(p_j,q) A_n(\Lambda_1^+,g_2^+,...,\Lambda_j^-,...,g_{n-1}^+,g_n^- ) 
 \nonumber \\
 & &  
       - \Gamma^-(p_n,q) A_n(\Lambda_1^+,g_2^+,...,g_j^-,...,g_{n-1}^+, \Lambda_n^-).
\eq
If one further sets 
the reference momentum equal to $q=p_j$ and uses the expression 
for the maximally helicity violating gluon amplitudes 
one obtains the expression for an amplitude with a pair of quarks:
\bq
 A_{n}(q_1^+,g_2^+,...,g_j^-,...,g_{n-1}^+,\bar{q}_n^-) 
 & = & i \left( \sqrt{2} \right)^{n-2}
 \frac{\langle j 1 \rangle \langle j n \rangle^3}
      {\langle 1 2 \rangle \langle 2 3 \rangle ... \langle n 1 \rangle}.
\eq


\subsection{New developments: Twistor methods}
\label{twistors}

In the previous section we saw that for the all-gluon tree amplitude there are remarkable simple
formul{\ae} if almost all gluons have one helicity and not more than two gluons have the 
opposite helicity.
Of course we are also interested in the case, where more than two gluons have the opposite helicity.
MHV vertices, discussed in section \ref{sect:mhv_vertices}, 
provide an answer to this question and tell us how the complexity increases
with the number of opposite helicity gluons.
In addition this construction led to new recursion relation, which no longer require that one external
leg is kept off-shell.
The building blocks of these on-shell recursion relations are gauge-invariant amplitudes.
This is discussed in section \ref{sect:on_shell}.

\subsubsection{MHV vertices}
\label{sect:mhv_vertices}

As an alternative to usual Feynman graphs, tree amplitudes in Yang-Mills theory
can be constructed from tree graphs in which the vertices are tree level MHV
scattering amplitudes, continued off shell in a particular fashion \cite{Cachazo:2004kj}.
The basic building blocks are the MHV amplitudes, which serve as new vertices:
\bq
\label{mhv_vertex}
V_n(1^+,...,j^-,...,k^-,...,n^+) 
 & = & i \left( \sqrt{2} \right)^{n-2} 
 \frac{\langle j k \rangle^4}{\langle 1 2 \rangle ... \langle n 1 \rangle}.
\eq
Each MHV vertex has exactly two lines carrying negative helicity and at least one line carrying positive helicity.
Each internal line has a positive helicity label on one side and a negative helicity label on the other side.
The propagator for each internal line is the propagator of a scalar particle:
\bq
 \frac{i}{k^2}
\eq
The expression~(\ref{mhv_vertex}) for the MHV vertices involves spinors corresponding to massless on-shell momenta $k_j^2=0$.
Therefore we have state what this light-like four-vector should be for every internal line meeting a MHV vertex.
As in eq.~(\ref{projection_null}) the light-like four-vector can be taken as \cite{Kosower:2004yz}
\bq
\label{projection_null_2}
 k^\flat & = & k - \frac{k^2}{2 k \cdot q} q,
\eq
where $k$ is the momentum flowing through the internal line and $q$ is a fixed light-like reference momentum.
\begin{figure}
\begin{center}
\includegraphics[bb= 95 575 490 715,width=0.8\textwidth]{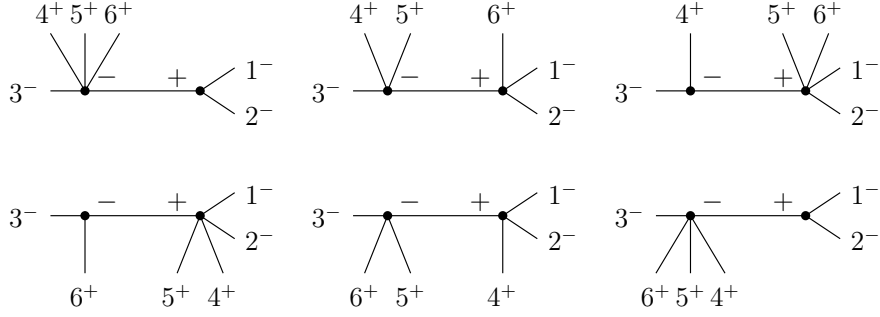}
\end{center}
\caption{MHV diagrams contributing to the tree-level six-gluon amplitude $A_6(1^-,2^-,3^-,4^+,5^+,6^+)$.}
\label{fig5}
\end{figure}
Let us now consider an example. The amplitude $A_6(1^-,2^-,3^-,4^+,5^+,6^+)$ has three gluons of positive helicity
and three gluons of negative helicity and is one of the first non-trivial amplitudes, which are non-zero and which are not
MHV amplitudes.
Fig. \ref{fig5} shows the six MHV diagrams contributing to this amplitude.
The first diagram yields
\bq
\lefteqn{
\begin{picture}(120,60)(0,40)
 \Vertex(25,50){2}
 \Vertex(75,50){2}
 \Line(25,50)(75,50)
 \Line(10,50)(25,50)
 \Line(75,50)(90,60)
 \Line(75,50)(90,40)
 \Text(95,60)[l]{\scriptsize $1^-$}
 \Text(95,40)[l]{\scriptsize $2^-$}
 \Text(5,50)[r]{\scriptsize $3^-$}
 \Text(30,52)[lb]{\scriptsize $-$}
 \Text(70,52)[rb]{\scriptsize $+$}
 \Line(25,50)(10,75)
 \Line(25,50)(25,75)
 \Line(25,50)(40,75)
 \Text(10,80)[b]{\scriptsize $4^+$}
 \Text(25,80)[b]{\scriptsize $5^+$}
 \Text(40,80)[b]{\scriptsize $6^+$}
\end{picture}
 = }
 \nonumber \\
 & & \nonumber \\
 & &
 \left[
 i \sqrt{2} 
 \frac{\langle 1 2 \rangle^4}{\langle 1 2 \rangle \langle 2 \left(-k_{12}^\flat\right) \rangle
 \langle \left(-k_{12}^\flat\right) 1 \rangle}
 \right]
 \;\;\;
 \frac{i}{k_{12}^2}
 \;\;\;
 \left[
 i \left( \sqrt{2} \right)^{3} 
 \frac{\langle 3 k_{12}^\flat \rangle^4}{\langle 3 4 \rangle 
 \langle 4 5\rangle \langle 5 6 \rangle 
 \langle 6 k_{12}^\flat \rangle \langle k_{12}^\flat 3 \rangle }
 \right],
\eq
and similar expressions are obtained for the five other diagrams.
In this expression, $k_{12}=k_1+k_2$ is the momentum flowing through the internal line and $k_{12}^\flat$ is the
projection onto a light-like four-vector as in eq.~(\ref{projection_null_2}).
We recall from table \ref{table2} that a brute force approach would require the calculation of 220 Feynman diagrams.
Restricting ourselves to a partial amplitude with a fixed cyclic order reduces this number to 36 diagrams.
In the approach based on MHV vertices there are only six diagrams.
In the next subsection we will discuss a method which reduces the number of diagrams even further.

\subsubsection{On-shell recursion relations}
\label{sect:on_shell}

Britto, Cachazo and Feng \cite{Britto:2004ap}
gave a recursion relation for the calculation of the $n$-gluon amplitude
which involves only on-shell amplitudes.
To describe this method it is best not to view the partial amplitude $A_n$ as a function
of the four-momenta $k_j^\mu$, but to replace each four-vector by a pair of two-component
Weyl spinors.
In detail this is done as follows:
Each four-vector $K_\mu$ has a bi-spinor representation, given by
\bq
\label{bispinor_representation}
 K_{A\dot{B}} = K_\mu \sigma^\mu_{A\dot{B}},
 & &
 K_\mu = \frac{1}{2} K_{A\dot{B}} \bar{\sigma}_\mu^{\dot{B}A}.
\eq
For light-like vectors this bi-spinor representation factorises into a dyad of Weyl spinors:
\bq
\label{dyad}
 k_\mu k^\mu = 0
 & \Leftrightarrow &
 k_{A\dot{B}} = k_{A} k_{\dot{B}}.
\eq
The equations~(\ref{bispinor_representation}) and~(\ref{dyad})
allow us to convert any light-like four-vector into a dyad of Weyl spinors and vice versa.
Therefore the partial amplitude $A_n$, being originally a function of the momenta $k_j$ and helicities
$\lambda_j$,
can equally be viewed as a function of the Weyl spinors $k_A^j$, $k_{\dot{B}}^j$ and the helicities
$\lambda_j$:
\bq
 A_n(k_1^{\lambda_1},...,k_n^{\lambda_n}) & = & 
 A_n( k_A^1, k_{\dot{B}}^1, \lambda_1, ..., k_A^n, k_{\dot{B}}^n, \lambda_n).
\eq
Note that for an arbitrary pair of Weyl spinors, the corresponding four-vector will in general be complex-valued.
\begin{figure}
\begin{center}
\includegraphics[bb= 90 610 560 720,width=0.8\textwidth]{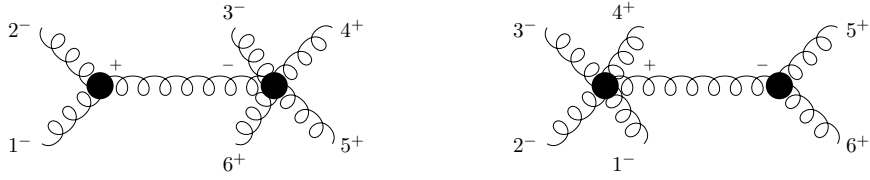}
\end{center}
\caption{Diagrams contributing to the tree-level six-gluon amplitude $A_6(1^-,2^-,3^-,4^+,5^+,6^+)$ in the on-shell 
recursive approach. The vertices are on-shell amplitudes.}
\label{fig6}
\end{figure}
If $(\lambda_n,\lambda_1) \neq (+,-)$ we have the following recurrence relation:
\bq
\label{on_shell_recursion}
\lefteqn{
A_n\left( k_A^1, k_{\dot{B}}^1, \lambda_1, ..., k_A^n, k_{\dot{B}}^n, \lambda_n\right)
 = 
 } & & \\
 & & 
 \hspace*{20mm}
 \sum\limits_{j=3}^{n-1} \sum\limits_{\lambda=\pm}
  A_{j}\left( \hat{k}_A^1, k_{\dot{B}}^1, \lambda_1, 
              k_A^2, k_{\dot{B}}^2, \lambda_2, 
              ..., 
	      k_A^{j-1}, k_{\dot{B}}^{j-1}, \lambda_{j-1},
              i \hat{K}_A, i \hat{K}_{\dot{B}}, -\lambda
              \right)
 \nonumber \\
 & &
 \hspace*{20mm}
  \times
  \frac{i}{K_{1,j-1}^2} 
  A_{n-j+2}\left( 
                  \hat{K}_A, \hat{K}_{\dot{B}}, \lambda,
                  k_A^j, k_{\dot{B}}^j, \lambda_j, 
                  ..., 
                  k_A^{n-1}, k_{\dot{B}}^{n-1}, \lambda_{n-1},
                  k_A^n, \hat{k}_{\dot{B}}^n, \lambda_n \right).
 \nonumber 
\eq
If $(\lambda_n,\lambda_1) = (+,-)$ we can always cyclic permute the arguments, such that
$(\lambda_n,\lambda_1) \neq (+,-)$.
This is possible, since on-shell amplitudes, where all gluons have the same helicity, vanish.
In eq.~(\ref{on_shell_recursion}) the shifted spinors
$\hat{k}_A^1$, $\hat{k}_{\dot{B}}^n$, $\hat{K}_A$ and $\hat{K}_{\dot{B}}$ are given by
\bq 
 \hat{k}_A^1 = k_A^1 - z k_A^n, 
 & &
 \hat{K}_A = \frac{K_{A\dot{B}} k_1^{\dot{B}}}{\sqrt{\left\l 1+ \left| K \right| n+ \right\r}},
 \nonumber \\
 \hat{k}_{\dot{B}}^n = k_{\dot{B}}^n + z k_{\dot{B}}^1,
 & &
 \hat{K}_{\dot{B}} = \frac{k_n^A K_{A\dot{B}}}{\sqrt{\left\l 1+ \left| K \right| n+ \right\r}},
\eq
where
\bq
 K_{A\dot{B}} = \sum\limits_{l=1}^{j-1} k_A^l k_{\dot{B}}^l, 
 & & 
 K_{1,j-1}^2 = \mbox{det} \; K_{A\dot{B}},
 \;\;\; \mbox{and}\;\;\;
 z = \frac{K_{1,j-1}^2}{\left\l 1+ \left| K \right| n+ \right\r}.
\eq
Let us again consider as an example the amplitude $A_6(1^-,2^-,3^-,4^+,5^+,6^+)$.
The diagrams contributing in the on-shell approach are shown in fig. \ref{fig6}.
One obtains as a result for the amplitude
\bq
\lefteqn{
A_6(1^-,2^-,3^-,4^+,5^+,6^+) = } & & 
 \nonumber \\
 & &
 4 i \left[
           \frac{\l 6+ | 1+2 | 3+ \r^3}{[ 61 ] [ 12 ] \l 34 \r \l 45 \r s_{126} \l 2+ | 1+6 | 5+ \r}
         + \frac{\l 4+ | 5+6 | 1+ \r^3}{[ 23 ] [ 34 ] \l 56 \r \l 61 \r s_{156} \l 2+ | 1+6 | 5+ \r}
     \right].
\eq
Note that there only two diagrams, which need to be calculated.
\begin{table}
\begin{center}
\begin{tabular}{l|r}
method & diagrams \\
\hline
brute force approach &  220 \\
colour-ordered amplitudes & 36 \\
MHV vertices & 6 \\
on-shell recursion & 2 \\
\end{tabular}
\end{center}
\caption{The number of diagrams contributing to the colour-ordered 
six-gluon amplitude $A_6(1^-,2^-,3^-,4^+,5^+,6^+)$ using various methods.}
\label{table3}
\end{table}
Table \ref{table3} shows a comparison of the number of diagrams
contributing to the colour-ordered 
six-gluon amplitude in the various approaches.
The performance of a numerical implementation of these new methods have been 
investigated in \cite{Dinsdale:2006sq,Duhr:2006iq}.

\subsubsection{Proof of the on-shell recursion relations}
\label{sect:on_shell_proof}

For the proof \cite{Britto:2005fq,Badger:2005zh,Risager:2005vk,Draggiotis:2005wq,Vaman:2005dt,Schwinn:2007ee}
of the on-shell recursion relation one considers the function
\bq
A(z) & = &
A_n\left( \hat{k}_A^1, k_{\dot{B}}^1, \lambda_1, ..., k_A^n, \hat{k}_{\dot{B}}^n, \lambda_n\right)
\eq
of one variable $z$,
where the $z$-dependence enters through 
\bq 
 \hat{k}_A^1 = k_A^1 - z k_A^n, 
 & &
 \hat{k}_{\dot{B}}^n = k_{\dot{B}}^n + z k_{\dot{B}}^1.
\eq
Note that for all values of $z$, this is an on-shell amplitude. 
However, the four-momenta of particles $1$ and $n$ are in general complex.
The function $A(z)$ is a rational function of $z$, which has only simple poles in $z$.
This follows from the Feynman rules and the factorisation properties of amplitudes.
Therefore, if $A(z)$ vanishes for $z \rightarrow \infty$, $A(z)$ is given 
by Cauchy's theorem as the sum over its residues. This is just the right hand side of the recursion relation.
The essential ingredient for the proof is the vanishing of $A(z)$ at $z \rightarrow \infty$.
If $(\lambda_1,\lambda_n)=(+,-)$ it can be shown that each individual Feynman diagram 
vanishes for $z \rightarrow \infty$. Consider the flow of the $z$-dependence in a particular diagram
The most dangerous contribution comes from a path, where all vertices are
three-gluon-vertices. For a path made of $n$ propagators we have $n+1$ vertices and the product of propagators
and vertices behaves therefore like $z$ for large $z$. 
This statement remains true for a path containing only one vertex and no propagators.
The polarisation vectors for the
helicity combination $(\lambda_1,\lambda_n)=(+,-)$ contribute a factor $1/z^2$, therefore the complete diagram
behaves like $1/z$ and vanishes therefore for $z \rightarrow \infty$.


\section{Calculating loop amplitudes}
\label{sect:loop}

The second bottle-neck for higher-order computations are loop integrals.
I first discuss one-loop integrals in section \ref{sect:one_loop}. These are relevant 
for multi-leg NLO calculations, like the processes listed in table \ref{table1}.
In section \ref{sect:two_loop} I discuss techniques for two-loop amplitudes and beyond.

\subsection{Automated computation of one-loop amplitudes}
\label{sect:one_loop}

The simplest, but most important loop integrals are the one-loop integrals.
We have a good understanding of these integrals and I will present the main results in this section.
An important result is that any scalar integral with more than four external legs can be reduced to 
scalar integrals with no more than $4$ external legs.
Therefore the set of basic one-loop integrals is rather limited. I will discuss this reduction in
section \ref{subsect:reduction_higher_point}.
For one-loop tensor integrals we can use the Passarino-Veltman method, which reduces any tensor
integral to a combination of scalar integrals.
Improvements of the Passarino-Veltman algorithm are discussed in section \ref{sect:passarino}.
Section \ref{sect:unitarity} is devoted to methods, which avoid Feynman diagrams.

\subsubsection{Reduction to integrals with no more than four external legs}
\label{subsect:reduction_higher_point}

In this section we discuss the reduction of scalar integrals with more than four external legs
to a basic set of scalar one-, two-, three- and four-point functions.
It is a long known fact, that higher point scalar integrals can be expressed
in terms of this basic set \cite{Melrose:1965kb,vanNeerven:1984vr}, however
the practical implementation within dimensional regularisation
was only worked out 
recently \cite{Bern:1994kr,Binoth:1999sp,Fleischer:1999hq,Denner:2002ii,Duplancic:2003tv}.
The one-loop $n$-point functions with $n \ge 5$ are always UV-finite, but they may have IR-divergences.
Let us first assume that there are no IR-divergences. Then the integral is finite and can be performed 
in four dimensions. In a space of four dimensions we can have no more than four linearly independent vectors,
therefore it comes to no surprise that in an one-loop integral with five or more propagators, one propagator
can be expressed through the remaining ones. This is the basic idea for the reduction of 
the higher point scalar integrals.
With slight modifications it can be generalised to dimensional regularisation.
I will discuss the method for massless one-loop integrals
\bq
\label{basic_one_loop_scalar_int}
I_n & = &
 e^{\eps \gamma_E} \mu^{2\eps} (-1)^n
  \int \frac{d^Dk}{i \pi^{\frac{D}{2}}}
  \frac{1}{k^2 (k-p_1)^2 ... (k-p_1-...p_{n-1})^2}.
\eq
With the notation
\bq
 q_i & = & \sum\limits_{j=1}^i p_j
\eq
one can associate two matrices $S$ and $G$ to the integral in eq.~(\ref{basic_one_loop_scalar_int}).
The entries of the $n \times n$ kinematical matrix $S$ are given by
\bq
 S_{ij} & = & \left( q_i - q_j \right)^2,
\eq
and the entries of the $(n-1) \times (n-1)$ Gram matrix are defined by
\bq
G_{ij} & = & 2 q_i q_j.
\eq
For the reduction one distinguishes three different cases: 
Scalar pentagons (i.e. scalar five-point functions),
scalar hexagons (scalar six-point functions) and scalar integrals with more than
six propagators.

Let us start with the pentagon.
A five-point function in $D=4-2\eps$ dimensions can be expressed as a sum of four-point functions, where
one propagator is removed, plus a five-point function in $6-2\eps$ dimensions \cite{Bern:1994kr}.
Since the $(6-2\eps)$-dimensional pentagon is finite and comes with an extra factor of $\eps$ in front, it does
not contribute at $O(\eps^0)$. In detail we have
\bq
I_5 & = & -2\eps B I_5^{6-2\eps}
          - \sum\limits_{i=1}^5 b_i I_4^{(i)}
 =
          - \sum\limits_{i=1}^5 b_i I_4^{(i)}
  + O\left(\eps\right),
\eq
where $I_5^{6-2\eps}$ denotes the $(6-2\eps)$-dimensional pentagon and
$I_4^{(i)}$ denotes the four-point function, which is obtained from the pentagon by removing propagator $i$.
The coefficients $B$ and $b_i$ are obtained from the kinematical matrix $S_{ij}$ as follows: 
\bq
b_i = \sum\limits_j \left( S^{-1} \right)_{ij},
 & &
B = \sum\limits_{i} b_i.
\eq
The six-point function can be expressed as a sum of five-point functions \cite{Binoth:1999sp}
without any correction of $O(\eps)$
\bq
\label{scalarsixpoint}
I_6 & = & - \sum\limits_{i=1}^6 b_i I_5^{(i)},
\;\;\;
b_i = \sum\limits_j \left( S^{-1} \right)_{ij},
\eq
where the coefficients $b_i$ are again related to the kinematical matrix $S_{ij}$.
For the seven-point function and beyond we can again express the $n$-point function as a sum over
$(n-1)$-point functions \cite{Duplancic:2003tv}: 
\bq
\label{scalarnpoint}
I_n & = & - \sum\limits_{i=1}^n r_i I_{n-1}^{(i)}.
\eq
In contrast to eq. (\ref{scalarsixpoint}), the decomposition in eq. (\ref{scalarnpoint}) is no longer unique.
A possible set of coefficients $r_i$ can be
obtained from the singular value decomposition of the Gram matrix 
\bq
G_{ij} & = & \sum\limits_{k=1}^4 U_{ik} w_k \left(V^T\right)_{kj}.
\eq
as follows \cite{Giele:2004iy}
\bq
  r_i = \frac{V_{i 5}}{W_5}, \;\;\; 1 \le i \le n-1,
  \;\;\; \;\;\; 
  r_n = - \sum\limits_{j=1}^{n-1} r_j,
 \;\;\; \;\;\; 
 W_5 = \frac{1}{2} \sum\limits_{j=1}^{n-1} G_{j j} V_{j 5}.
\eq

\subsubsection{Improvements of the Passarino-Veltman algorithm}
\label{sect:passarino}

We now consider the reduction of tensor loop integrals 
(e.g. integrals, where the loop momentum appears in the numerator)
to a set of scalar loop integrals (e.g. integrals, where the numerator
is independent of the loop momentum).
For one-loop integrals a systematic algorithm has been first worked
out by Passarino and Veltman \cite{Passarino:1979jh}.
Consider the following three-point integral
\bq
I_3^{\mu\nu} & = & \int \frac{d^{D}k}{i \pi^{D/2}}
\frac{k^\mu k^\nu}{k^2 (k-p_1)^2 (k-p_1-p_2)^2},
\eq
where $p_1$ and $p_2$ denote the external momenta.
The reduction technique according to Passarino and Veltman consists in writing $I_3^{\mu\nu}$
in the most general form
in terms of form factors times external momenta and/or the metric tensor. In our example above
we would write
\bq
\label{passarino}
I_3^{\mu\nu} & = & p_1^\mu p_1^\nu C_{21} + p_2^\mu p_2^\nu C_{22}
+ \{p_1^\mu, p_2^\nu\} C_{23} + g^{\mu\nu} C_{24},
\eq
where $\{p_1^\mu, p_2^\nu\} = p_1^\mu p_2^\nu + p_2^\mu p_1^\nu$.
One then solves for the form factors $C_{21}, C_{22}, C_{23}$ and $C_{24}$ by
first contracting both sides with the external momenta $p_1^\mu p_1^\nu$, $p_2^\mu p_2^\nu$,
$\{p_1^\mu, p_2^\nu \}$ and the metric tensor $g^{\mu\nu}$.
On the left-hand side the resulting scalar products between the loop momentum $k^\mu$ and the external
momenta are rewritten in terms of the propagators, as for example
\bq
2 p_1 \cdot k & = & k^2 - (k-p_1)^2 + p_1^2.
\eq
The first two terms of the right-hand side above cancel propagators, whereas the last term does not involve the 
loop momentum anymore.
The remaining step is to solve for the form-factors $C_{2i}$ by inverting the matrix which one obtains on the 
right-hand side of
equation (\ref{passarino}).
Due to this step Gram determinants usually appear in the denominator of the final expression.
In the example above we would encounter the Gram determinant of the triangle
\bq
\Delta_3 & = & 4 \left|
\begin{array}{cc}
p_1^2 & p_1\cdot p_2 \\
p_1 \cdot p_2 & p_2^2 \\
\end{array} \right|.
\eq
One drawback of this algorithm is closely related to these determinants : In a phase space
region where $p_1$ becomes collinear to $p_2$, the Gram determinant will tend to zero, and the 
form factors will take large values, with possible large cancellations among them. This makes
it difficult to set up a stable numerical program for automated evaluation of tensor loop
integrals.
Quite some effort went therefore into improvements and alternatives, which avoid these instabilities.
They are centered around the following ideas:
\begin{itemize}
\item The use of a different reduction scheme in critical regions, based on an 
expansion around the small invariants \cite{Denner:2005nn,Dittmaier:2003bc,Stuart:1990de,Stuart:1988tt}.
\item Choosing a different set of basic integrals. The new set can be non-minimal. Quite often the new
basic integrals correspond to scalar integrals in higher dimensions.
\cite{Davydychev:1991va,Bern:1992em,Bern:1993kr,Campbell:1997zw,Giele:2004iy,Giele:2004ub,Ellis:2005zh}.
\item Direct numerical integration \cite{Ferroglia:2002mz,Kramer:2002cd,Nagy:2003qn,Binoth:2002xh,Binoth:2005ff,Anastasiou:2007qb}.
\item The use of spinor techniques, which avoid to a certain extent the occurrence of Gram determinants \cite{Pittau:1997ez,Pittau:1998mv,Weinzierl:1998we,del Aguila:2004nf,van Hameren:2005ed,Ossola:2006us}.
\end{itemize}

\subsubsection{Methods based on unitarity}
\label{sect:unitarity}

We know that the final answer for a one-loop amplitude in a massless gauge theory
can be written as
\bq
\label{cutbased}
A_n^{(1)} & = & 
 \sum\limits_i c_i I_2^{(i)} 
+ \sum\limits_{i,j} c_{ij} I_3^{(ij)}
+ \sum\limits_{i,j,k} c_{ijk} I_4^{(ijk)}
+ R.
\eq
$I_2$, $I_3$ and $I_4$ are the scalar bubble, triangle and  box integral functions.
In a massive theory we would have in addition also scalar one-point functions.
\begin{figure}
\begin{center}
\includegraphics[bb= 60 500 480 605,width=0.8\textwidth]{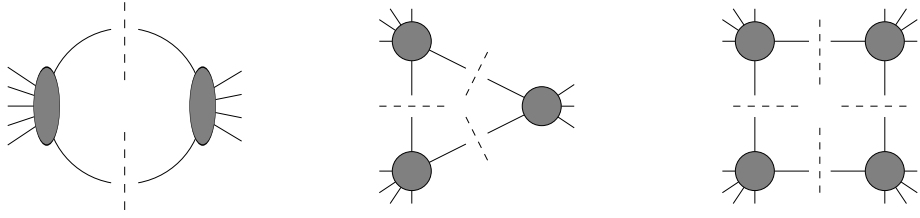}
\end{center}
\caption{Double, triple and quadruple cuts.}
\label{fig10}
\end{figure}
In a massless theory these functions are zero within dimensional regularisation.
Note that there are no integral functions with more than four internal propagators.
These higher-point functions can always be reduced to the set above, as we have seen
in section \ref{subsect:reduction_higher_point}.
$R$ is called the rational term.
The set of all occurring integral functions
\bq
 {\cal F} & = & \{ I_2^{(i)}, I_3^{(ij)}, I_4^{(ijk)} \}
\eq
is rather easily obtained from pinching in all possible ways internal propagators in all occurring diagrams.
We can assume that we know this set in advance.
To compute the amplitude requires therefore the determination of the coefficients
$c_i$, $c_{i,j}$, $c_{i,j,k}$ and the rational term $R$.
In eq.~(\ref{cutbased}) any $\eps$-dependence of the coefficients has been removed. The original $\eps$-dependent
parts of the coefficients, which are multiplied with poles from the integral functions 
are collected in the rational term $R$.
Within the unitarity-based methods one calculates the coefficients and the rational term without resorting
to a Feynman diagram calculation.

The original formulation of the method \cite{Bern:1994zx,Bern:1995cg} is based on the observation 
that the basic integral functions contain logarithms and dilogarithms, which
develop imaginary parts in certain regions
of phase space, for example
\bq
\mbox{Im} \; \ln \left( \frac{-s-i 0}{-t-i 0} \right) & = & - \pi \left[ \theta(s) -\theta(t) \right], \nonumber \\
\mbox{Im} \; \mbox{Li}_2 \left( 1 - \frac{(-s-i 0)}{(-t-i 0)} \right)
 & = & - \ln \left( 1 - \frac{s}{t} \right)
        \mbox{Im} \; \ln \left( \frac{-s-i 0}{-t-i 0} \right).
\eq
Knowing the imaginary parts, one can reconstruct uniquely the corresponding
integral functions.
In general there will be imaginary parts corresponding to different
channels (e.g. to the different possibilities to cut a one-loop
diagram into two parts).
The imaginary part in one channel of a one-loop amplitude 
can be obtained via unitarity from a phase space integral over
two tree-level amplitudes.
With the help of the Cutkosky rules we have
\bq
\label{cutconstr}
\mbox{Im} \; A^{(1)} & = & \mbox{Im} \int \frac{d^D k}{(2 \pi)^D} \frac{1}{k_1^2}
\frac{1}{k_2^2} A_L^{(0)} A_R^{(0)}.
\eq
$A^{(1)}$ is the one-loop amplitude under consideration, $A^{(0)}_L$ and
$A^{(0)}_R$ are tree-level amplitudes appearing on the left and right side
of the cut in a given channel, as shown in the first picture of fig. \ref{fig10}.
Lifting eq. (\ref{cutconstr}) one obtains
\bq
A^{(1)} & = & \int \frac{d^D k}{(2 \pi)^D} \frac{1}{k_1^2}
\frac{1}{k_2^2} A_L^{(0)} A_R^{(0)} + \;\mbox{cut free pieces},
\eq
where ``cut free pieces'' denote contributions which do not develop an imaginary
part in this particular channel.
By evaluating the cut, one determines the coefficients of the integral
functions, which have an imaginary part in this channel.
Iterating over all possible cuts, one finds all coefficients.
One advantage of a cut-based calculation is that one starts with tree amplitudes on both sides of the cut, which are already sums
of Feynman diagrams. 
Therefore cancellations and simplifications, which usually occur between
various diagrams, can already be performed before we start the calculation
of the loop amplitude.
This technique was used in the calculation of the one-loop amplitudes for $e^+ e^- \rightarrow \mbox{4 partons}$
\cite{Bern:1997ka,Bern:1997sc}.
The rational part $R$ can be obtained by calculating higher order
terms in $\eps$ within the cut-based method.
At one-loop order an arbitrary scale $\mu^{2\varepsilon}$ is introduced in order to keep the coupling
dimensionless. In a massless theory the factor $\mu^{2\varepsilon}$ is always accompanied
by some kinematical invariant $s^{-\varepsilon}$ for dimensional reasons.
If we write symbolically
\bq
A^{loop} & = & \frac{1}{\varepsilon^2} c_2 \left( \frac{s_2}{\mu^2} \right)^{-\varepsilon} 
+ \frac{1}{\varepsilon} c_1 \left( \frac{s_1}{\mu^2} \right)^{-\varepsilon}
+ c_0 \left( \frac{s_0}{\mu^2} \right)^{-\varepsilon} ,
\eq
the cut-free pieces $c_0 (s_0/\mu^2)^{-\varepsilon}$ can be detected at order $\varepsilon$:
\bq
c_0 \left( \frac{s_0}{\mu^2} \right)^{-\varepsilon} & = & c_0 - \varepsilon c_0 \ln \left(\frac{s_0}{\mu^2}\right) + O(\varepsilon^2).
\eq
With the advent of the new methods based on twistors, significant improvements were added to the unitarity-based
technique \cite{Britto:2004nc,Bidder:2004tx,Bidder:2004vx,Bidder:2005ri,Bedford:2004nh,Britto:2005ha,Bern:2005hs,Bern:2005ji,Bern:2005cq,Forde:2005hh,Berger:2006ci,Berger:2006vq,Britto:2006sj,Anastasiou:2006jv,Anastasiou:2006gt,Mastrolia:2006ki,Britto:2006fc,Forde:2007mi}.
Apart from the two-particle cut discussed above, one can also consider triple or quadruple cuts
as shown in fig. \ref{fig10}.
A particular nice result follows from quadruple cuts \cite{Britto:2004nc}: The coefficients of the box integral functions
are given as a product of four tree amplitudes, summed over the two solutions of the on-shell conditions
\bq
 l_1^2 = l_2^2 = l_3^2 = l_4^2 = 0.
\eq
Unitarity-based methods contributed significantly to the calculation of the 
one-loop six-gluon amplitude \cite{Bern:1994zx,Bern:1995cg,Bidder:2004tx,Bidder:2004vx,Bidder:2005ri,Bedford:2004nh,Britto:2005ha,Bern:2005cq,Bern:2005hh,Britto:2006sj,Berger:2006ci,Berger:2006vq,Xiao:2006vr,Su:2006vs,Xiao:2006vt,Ellis:2006ss}
and the one-loop six-photon amplitude \cite{Binoth:2006hk,Nagy:2006xy,Binoth:2007ca,Ossola:2007bb,Forde:2007mi}.


\subsection{Two-loop amplitudes and beyond}
\label{sect:two_loop}

For NNLO calculation two-loop amplitudes are required. The relevant two-loop integrals are
far from trivial. In this section I review a few techniques which can used for the calculation
of two-loop integrals.
The Mellin-Barnes transformation is discussed in section \ref{sect:mellin_barnes}.
Multiple polylogarithms together with their algebraic properties are introduced in section \ref{sect:polylog}.
Section \ref{sect:decomp} is devoted to sector decomposition.
Not treated in detail, but equally important are methods which reduce the work-load:
Integration-by-parts identities \cite{Chetyrkin:1981qh} have a long-standing tradition.
In addition there are the reduction algorithms of Tarasov \cite{Tarasov:1996br,Tarasov:1997kx}
and Laporta \cite{Laporta:2001dd}, which can be used to relate all tensor integrals
to a set of basic scalar integrals.
With the help of these techniques many two-loop amplitudes have been calculated:
Bhabha scattering \cite{Bern:2000ie},
$p p \rightarrow \mbox{2 jets}$ \cite{Bern:2000dn,Anastasiou:2000kg,Anastasiou:2000ue,Anastasiou:2000mv,Anastasiou:2001sv,Glover:2001af,Bern:2002tk},
$e^+ e^- \rightarrow \mbox{3 jets}$ \cite{Garland:2001tf,Garland:2002ak,Moch:2002hm} and
Higgs production \cite{Harlander:2000mg,Ravindran:2004mb}.
In addition, the
three-loop splitting functions, required for the evolution of parton distribution functions at NNLO,
have been calculated \cite{Moch:2004pa,Vogt:2004mw}.

\subsubsection{The Mellin-Barnes transformation}
\label{sect:mellin_barnes}

The Mellin-Barnes transformation can be used to transform any Feynman parameter
integral into a particular simple form, such that the integrals over the Feynman parameters can 
be performed:
\bq
\label{multi_beta_fct}
 \int\limits_{0}^{1} \left( \prod\limits_{j=1}^{n}\,dx_j\,x_j^{\nu_j-1} \right)
 \delta(1-\sum_{i=1}^n x_i)
 & = & 
 \frac{\prod\limits_{j=1}^{n}\Gamma(\nu_j)}{\Gamma(\nu_1+...+\nu_n)}.
\eq
The Mellin-Barnes transformation reads
\bq
\lefteqn{
\left(A_1 + A_2 + ... + A_n \right)^{-c} 
 = 
 \frac{1}{\Gamma(c)} \frac{1}{\left(2\pi i\right)^{n-1}} 
 \int\limits_{-i\infty}^{i\infty} d\sigma_1 ... \int\limits_{-i\infty}^{i\infty} d\sigma_{n-1}
 } & & \\
 & & 
 \times 
 \Gamma(-\sigma_1) ... \Gamma(-\sigma_{n-1}) \Gamma(\sigma_1+...+\sigma_{n-1}+c)
 \; 
 A_1^{\sigma_1} ...  A_{n-1}^{\sigma_{n-1}} A_n^{-\sigma_1-...-\sigma_{n-1}-c}  
 \nonumber 
\eq
Each contour is such that the poles of $\Gamma(-\sigma)$ are to the right and the poles
of $\Gamma(\sigma+c)$ are to the left.
This transformation can be used to convert any Feynman parameter integral to the 
form of eq.~(\ref{multi_beta_fct}).
Therefore we exchange the Feynman parameter integrals against multiple complex contour integrals.
As this transformation converts sums into products it is 
the ``inverse'' of Feynman parametrisation.
The contour integrals are then performed by closing the contour at infinity and summing up all 
residues which lie inside the contour.
Here it is useful to know the residues of the Gamma function:
\bq
\mbox{res} \; \left( \Gamma(\sigma+a), \sigma=-a-n \right) = \frac{(-1)^n}{n!}, 
 & &
\mbox{res} \; \left( \Gamma(-\sigma+a), \sigma=a+n \right) = -\frac{(-1)^n}{n!}. 
\eq
Therefore we obtain (multiple) sum over residues. 
Techniques to manipulate these sums are discussed in the next section.
In particular simple cases the contour integrals can be performed in closed form with
the help of two lemmas of Barnes.
Barnes first lemma states that
\bq
\frac{1}{2\pi i} \int\limits_{-i\infty}^{i\infty} d\sigma
\Gamma(a+\sigma) \Gamma(b+\sigma) \Gamma(c-\sigma) \Gamma(d-\sigma) 
 =  
\frac{\Gamma(a+c) \Gamma(a+d) \Gamma(b+c) \Gamma(b+d)}{\Gamma(a+b+c+d)},
\;\;\;\;\;\;\;
\hspace*{-15mm}
\nonumber \\
\eq
if none of the poles of $\Gamma(a+\sigma) \Gamma(b+\sigma)$ coincides with the
ones from $\Gamma(c-\sigma) \Gamma(d-\sigma)$.
Barnes second lemma reads
\bq
\lefteqn{
\frac{1}{2\pi i} \int\limits_{-i\infty}^{i\infty} d\sigma
\frac{\Gamma(a+\sigma) \Gamma(b+\sigma) \Gamma(c+\sigma) \Gamma(d-\sigma) \Gamma(e-\sigma)}
{\Gamma(a+b+c+d+e+\sigma)} } & & \nonumber \\
& = & 
\frac{\Gamma(a+d) \Gamma(b+d) \Gamma(c+d) 
      \Gamma(a+e) \Gamma(b+e) \Gamma(c+e)}
{\Gamma(a+b+d+e) \Gamma(a+c+d+e) \Gamma(b+c+d+e)}.
\eq
Although the Mellin-Barnes transformation has been known for a long time, 
the method has seen a revival in applications in recent 
years \cite{Boos:1990rg,Davydychev:1990jt,Davydychev:1990cq,Smirnov:1999gc,Smirnov:1999wz,Tausk:1999vh,Smirnov:2000vy,Smirnov:2000ie,Smirnov:2003vi,Bierenbaum:2003ud,Heinrich:2004iq,Friot:2005cu,Bern:2005iz,Anastasiou:2005cb,Czakon:2005rk,Gluza:2007rt}.


\subsubsection{Multiple polylogarithms}
\label{sect:polylog}

The multiple polylogarithms are defined by
\bq 
\label{multipolylog2}
 \mbox{Li}_{m_1,...,m_k}(x_1,...,x_k)
  & = & \sum\limits_{i_1>i_2>\ldots>i_k>0}
     \frac{x_1^{i_1}}{{i_1}^{m_1}}\ldots \frac{x_k^{i_k}}{{i_k}^{m_k}}.
\eq
The multiple polylogarithms are generalisations of
the classical polylogarithms 
$
\mbox{Li}_n(x)
$ 
whose most prominent examples are
\bq
 \mbox{Li}_1(x) = \sum\limits_{i_1=1}^\infty \frac{x^{i_1}}{i_1} = -\ln(1-x),
 & &
 \mbox{Li}_2(x) = \sum\limits_{i_1=1}^\infty \frac{x^{i_1}}{i_1^2},
\eq 
as well as
Nielsen's generalised polylogarithms 
\bq
S_{n,p}(x) & = & \mbox{Li}_{n+1,1,...,1}(x,\underbrace{1,...,1}_{p-1}),
\eq
and the harmonic polylogarithms 
\bq
\label{harmpolylog}
H_{m_1,...,m_k}(x) & = & \mbox{Li}_{m_1,...,m_k}(x,\underbrace{1,...,1}_{k-1}).
\eq
Multiple polylogarithms and the closely related harmonic sums 
have been studied extensively in the literature
\cite{Hain,Goncharov,Borwein,Minh:2000,Remiddi:1999ew,Vermaseren:1998uu,Gehrmann:2000zt,Gehrmann:2001pz,Gehrmann:2001jv,Gehrmann:2002zr,Moch:2001zr,Blumlein:1998if,Blumlein:2003gb,Weinzierl:2004bn,Vollinga:2004sn,Korner:2005qz,Kalmykov:2006hu,Maitre:2007kp}.

In addition, multiple polylogarithms have an integral representation. 
To discuss the integral representation it is convenient to 
introduce for $z_k \neq 0$
the following functions
\bq
\label{Gfuncdef}
G(z_1,...,z_k;y) & = &
 \int\limits_0^y \frac{dt_1}{t_1-z_1}
 \int\limits_0^{t_1} \frac{dt_2}{t_2-z_2} ...
 \int\limits_0^{t_{k-1}} \frac{dt_k}{t_k-z_k}.
\eq
In this definition 
one variable is redundant due to the following scaling relation:
\bq
G(z_1,...,z_k;y) & = & G(x z_1, ..., x z_k; x y)
\eq
To relate the multiple polylogarithms to the functions $G$ it is convenient to introduce
the following short-hand notation:
\bq
\label{Gshorthand}
G_{m_1,...,m_k}(z_1,...,z_k;y)
 & = &
 G(\underbrace{0,...,0}_{m_1-1},z_1,...,z_{k-1},\underbrace{0...,0}_{m_k-1},z_k;y)
\eq
Here, all $z_j$ for $j=1,...,k$ are assumed to be non-zero.
One then finds
\bq
\label{Gintrepdef}
\mbox{Li}_{m_1,...,m_k}(x_1,...,x_k)
& = & (-1)^k 
 G_{m_1,...,m_k}\left( \frac{1}{x_1}, \frac{1}{x_1 x_2}, ..., \frac{1}{x_1...x_k};1 \right).
\eq
Eq. (\ref{Gintrepdef}) together with 
(\ref{Gshorthand}) and (\ref{Gfuncdef})
\begin{figure}
\begin{center}
\includegraphics[bb= 100 625 480 705,width=0.8\textwidth]{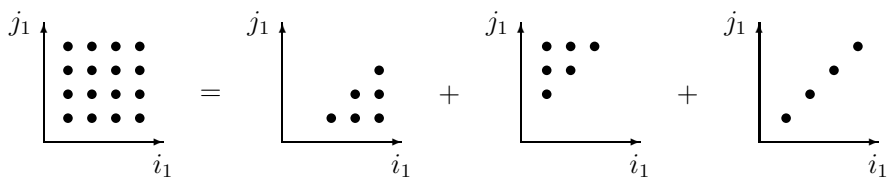}
\end{center}
\caption{Sketch of the proof for the quasi-shuffle product of nested sums. 
The sum over the square is replaced by
the sum over the three regions on the r.h.s.}
\label{fig8}
\end{figure}
defines an integral representation for the multiple polylogarithms.
Multiple polylogarithms form two (independent) algebras.
As a consequence, products of multiple polylogarithms can again be expressed as sums of
single polylogarithms.
To give an example, one has
\bq
\mbox{Li}_{m_1}(x_1) \mbox{Li}_{m_2}(x_2) 
& = & 
\mbox{Li}_{m_1,m_2}(x_1,x_2) + \mbox{Li}_{m_2,m_1}(x_2,x_1)
                                 + \mbox{Li}_{m_1+m_2}(x_1x_2),
 \nonumber \\
G(z_1;y) G(z_2;y) 
 & = & 
G(z_1,z_2;y) + G(z_2,z_1;y).
\eq
The first relation is based on the representation in terms of nested sums. A sketch of the proof
is shown in fig. \ref{fig8}.
\begin{figure}
\begin{center}
\includegraphics[bb= 120 645 460 725,width=0.8\textwidth]{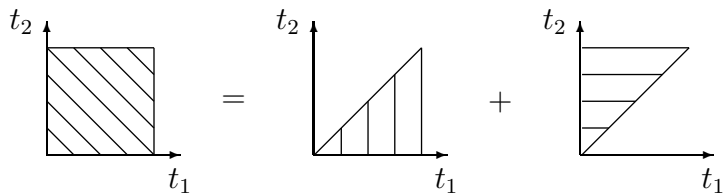}
\end{center}
\caption{Sketch of the proof for the shuffle product of two iterated integrals.
The integral over the square is replaced by two
integrals over the upper and lower triangle.}
\label{fig7}
\end{figure}
The second relation is based on the integral representation. The corresponding sketch of the proof
is shown in fig. \ref{fig7}.
These algebraic properties allow automated manipulations of expressions related to
multiple polylogarithms by computer algebra programs.
Several programs are available \cite{Vermaseren:1998uu,Weinzierl:2002hv,Moch:2005uc,Maitre:2005uu,Huber:2005yg}.

Up to now we treated multiple polylogarithms from an algebraic point of view.
Equally important are the analytical properties, which are needed for an efficient numerical 
evaluation.
As an example I first discuss the numerical evaluation of the dilogarithm \cite{'tHooft:1979xw}:
\bq
\mbox{Li}_{2}(x) & = & - \int\limits_{0}^{x} dt \frac{\ln(1-t)}{t}
 = \sum\limits_{n=1}^{\infty} \frac{x^{n}}{n^{2}}
\eq
The power series expansion can be evaluated numerically, provided $|x| < 1.$
Using the functional equations 
\bq
\mbox{Li}_2(x) & = & -\mbox{Li}_2\left(\frac{1}{x}\right) -\frac{\pi^2}{6} -\frac{1}{2} \left( \ln(-x) \right)^2,
 \nonumber \\
\mbox{Li}_2(x) & = & -\mbox{Li}_2(1-x) + \frac{\pi^2}{6} -\ln(x) \ln(1-x).
\eq
any argument of the dilogarithm can be mapped into the region
$|x| \le 1$ and
$-1 \leq \mbox{Re}(x) \leq 1/2$.
The numerical computation can be accelerated  by using an expansion in $[-\ln(1-x)]$ and the
Bernoulli numbers $B_i$:
\bq
\mbox{Li}_2(x) & = & \sum\limits_{i=0}^\infty \frac{B_i}{(i+1)!} \left( - \ln(1-x) \right)^{i+1}.
\eq
The generalisation for the numerical evaluation of multiple polylogarithms has been worked out in \cite{Vollinga:2004sn}.

\subsubsection{Sector decomposition}
\label{sect:decomp}

In this section I will discuss an algorithm, which allows to compute numerically the coefficients of the Laurent
expansion for a multi-loop integral for a given kinematical configuration of external momenta.
The major challenge such an algorithm has to face is the disentanglement of overlapping 
singularities.
An example for an overlapping singularity is given by
\bq
\label{example_sector_decomp}
     \int d^3 x \; \delta\left( 1 - \sum\limits_{i=1}^3 x_i \right) 
         \frac{x_1^{-1-\varepsilon} x_2^{-1-\varepsilon}}{x_1 + x_2}.
\eq
The term $1/(x_1+x_2)$ is an overlapping singularity. Sector decomposition 
\cite{Hepp:1966eg,Roth:1996pd,Binoth:2000ps}
is a convenient tool to disentangle
overlapping singularities. 
For the example in eq.~(\ref{example_sector_decomp})
one splits the integration region into two sectors $x_1>x_2$ and $x_1<x_2$.
\begin{figure}
\begin{center}
\includegraphics[bb= 100 560 480 725,width=0.6\textwidth]{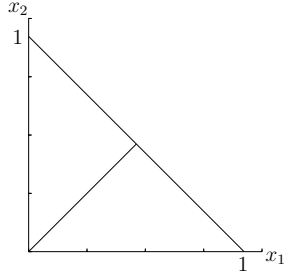}
\end{center}
\caption{Sector decomposition splits the integration over the large triangle $x_1+x_2<1$ into the sectors
$x_1>x_2$ and $x_2>x_1$.}
\label{fig9}
\end{figure}
This is shown in fig. \ref{fig9}
In the first sector one rescales $x_2$ as $x_2'=x_2/x_1$, while in the second sector one rescales
$x_1'=x_1/x_2$.
By applying the sector decomposition iteratively, 
one finally arrives at a form where all singularities are factorised 
explicitly in terms of factors of Feynman parameters like 
$x_j^{-1-\varepsilon}$. Subtractions of the form 
\bq
\int\limits_0^1 d x_j\,x_j^{-1-\varepsilon}\,f(x_j)
 & = & 
-\frac{1}{\varepsilon}\,\,f(0)
+\int\limits_0^1 d x_j\,x_j^{-1-\varepsilon}\,\left[ f(x_j)-f(0) \right]
\eq
for each $j$, where $\lim_{x_{j}\to 0}f(x_{j})$
is finite by construction, allow to extract all poles and lead to integrals which are finite and can
be integrated numerically.


\section{Cancellation of infrared divergences}
\label{sect:IR}

\begin{figure}
\begin{center}
\includegraphics[bb= 140 650 480 730,width=0.8\textwidth]{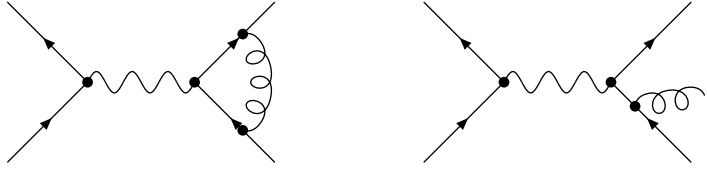}
\end{center}
\caption{Cancellation of infrared divergences between virtual corrections and real corrections.}
\label{fig11}
\end{figure}

Infrared divergences occur at next-to-leading order and beyond.
At NLO real and virtual corrections contribute.
The virtual corrections contain the loop integrals and can have,
in addition to ultraviolet divergences, infrared divergences.
If loop amplitudes are calculated in dimensional regularisation,
the IR divergences manifest themselves as
explicit poles in the 
dimensional regularisation parameter $\varepsilon=2-D/2$.
These poles cancel with similar poles arising from
amplitudes with additional partons but less internal loops, when integrated over phase space regions where
two (or more) partons become ``close'' to each other.
This is illustrated in fig. \ref{fig11}.
In general, the Kinoshita-Lee-Nauenberg theorem
\cite{Kinoshita:1962ur,Lee:1964is}
guarantees that any infrared-safe observable, when summed over all 
states degenerate according to some resolution criteria, will be finite.
However, the cancellation occurs only after the integration over the unresolved phase space
has been performed and prevents thus a naive Monte Carlo approach for a fully exclusive
calculation.
It is therefore necessary to cancel first analytically all infrared divergences and to use
Monte Carlo methods only after this step has been performed.

\subsection{Infrared divergences at NLO}

At NLO, general methods to circumvent this problem are known.
This is possible due to the universality of the singular behaviour
of the amplitudes in soft and collinear limits.
Examples are the phase-space slicing method
\cite{Giele:1992vf,Giele:1993dj,Keller:1998tf}
and the subtraction method
\cite{Frixione:1996ms,Catani:1997vz,Dittmaier:1999mb,Phaf:2001gc,Catani:2002hc}.
It is worth to examine a simple NLO example in detail to understand the basic concepts.
We consider the NLO corrections to $\gamma^\ast \rightarrow 2 \; \mbox{jets}.$
The real corrections are given by the matrix element for 
$\gamma^\ast \rightarrow q g \bar{q}$ and read, 
up to colour and coupling factors
\bq
 \left| {\cal A}_3 \right|^2 & = & 8 ( 1 - \varepsilon)  \left[
         \frac{2}{x_1 x_2} 
         - \frac{2}{x_1}
         - \frac{2}{x_2} 
         + (1-\varepsilon) \frac{x_2}{x_1}
         + (1-\varepsilon) \frac{x_1}{x_2} 
         - 2 \varepsilon 
        \right],
\eq
where $x_1=s_{12}/s_{123}$ and $x_2=s_{23}/s_{123}$.
This term is integrated over the three particle phase space.
Singularities occur at the boundaries of the integration region at $x_1=0$ and $x_2=0$.
Within the subtraction method 
one subtracts a suitable approximation term $d\sigma^A$ 
from the real corrections $d\sigma^R$.
This approximation term must have the same singularity structure as the real corrections.
If in addition the approximation term is simple enough, such that it can be integrated analytically
over a one-parton subspace, then the result can be added back to the virtual corrections $d\sigma^V$.
\bq
\sigma^{NLO} & = & \int\limits_{n+1} d\sigma^R + \int\limits_n d\sigma^V
= \int\limits_{n+1} \left( d\sigma^R - d\sigma^A \right) + \int\limits_n \left( d\sigma^V + \int\limits_1 
d\sigma^A \right).
\eq
Since by definition $d\sigma^A$ has the same singular behaviour as $d\sigma^R$, $d\sigma^A$
acts as a local counter-term and the combination $(d\sigma^R-d\sigma^A)$ is integrable
and can be evaluated numerically.
Secondly, the analytic integration of $d\sigma^A$ over the one-parton subspace will yield
the explicit poles in $\varepsilon$ needed to cancel the corresponding poles in $d\sigma^V$.
For the example discussed above the approximation term can be taken as a sum of two dipole subtraction
terms:
\bq
 & &
\left| {\cal A}_2(p_1',p_3') \right|^2 \frac{1}{s_{123}}
       \left[
         \frac{2}{x_1 (x_1 + x_2)} - \frac{2}{x_1}
         + (1-\varepsilon) \frac{x_2}{x_1}
        \right]
 \nonumber \\
 & &
 +
\left| {\cal A}_2(p_1'',p_3'') \right|^2 \frac{1}{s_{123}}
 \left[
         \frac{2}{x_2 (x_1 + x_2)} - \frac{2}{x_2}
         + (1-\varepsilon) \frac{x_1}{x_2}
        \right] 
\eq
The momenta $p_1'$, $p_3'$, $p_1''$ and $p_3''$ are linear combinations of the original momenta $p_1$, $p_2$ and $p_3$.
The first term is an approximation for $x_1 \rightarrow 0$, whereas the second term is an approximation
for $x_2 \rightarrow 0$.
Note that the soft singularity is shared between the two dipole terms 
and that in general the Born amplitudes ${\cal A}_2$ are evaluated with different momenta.
Antenna subtraction 
\cite{Kosower:1998zr,Daleo:2006xa}
allows to reduce the number of subtraction terms needed and 
interpolates smoothly between the $x_1 \rightarrow 0$ and
$x_2 \rightarrow 0$ regions.
Within antenna subtraction one could take for our example as approximation
\bq
\left| {\cal A}_2(p_1''',p_3''') \right|^2 \frac{1}{s_{123}}
       \left[
         \frac{2}{x_1 x_2} 
         - \frac{2}{x_1}
         - \frac{2}{x_2} 
         + (1-\varepsilon) \frac{x_2}{x_1}
         + (1-\varepsilon) \frac{x_1}{x_2} 
        \right].
\eq
Again, the Born amplitude ${\cal A}_2$ is evaluated with momenta $p_1'''$ and $p_2'''$, which are
linear combinations of the original momenta $p_1$, $p_2$ and $p_3$.
As a rule of thumb, one antenna subtraction term equals two dipol subtraction terms.

The dipole subtraction scheme has been worked out for general NLO calculations.
The matrix element corresponding to the approximation term $d\sigma^A$ is given as a sum over 
dipoles:
\bq
 \sum\limits_{pairs\; i,j} \;\;\; \sum\limits_{k \neq i,j} {\cal D}_{ij,k}.
\eq
Each dipole contribution has the following form:
\bq
{\cal D}_{ij,k} 
& = & 
{\cal A}_{n+2}^{(0)\;\ast}\left( p_1, ..., \tilde{p}_{(ij)},...,\tilde{p}_k,...\right)
\frac{(-{\bf T}_k \cdot {\bf T}_{ij})}{{\bf T}^2_{ij}} 
\frac{V_{ij,k}}{2 p_i \cdot p_j}
{\cal A}_{n+2}^{(0)}\left( p_1, ..., \tilde{p}_{(ij)},...,\tilde{p}_k,...\right).
 \;\;\;
\eq
Here ${\bf T}_i$ denotes the colour charge operator for parton $i$ and
$V_{ij,k}$ is a matrix in the spin space of the emitter parton $(ij)$.
In general, the operators ${\bf T}_i$ lead to colour correlations, while the $V_{ij,k}$'s lead
to spin correlations. The generation of all subtraction terms can be automated \cite{Weinzierl:2005dd}.


\subsection{Infrared divergences at NNLO}

The following terms contribute at NNLO:
\bq
d\sigma_{n+2}^{(0)} & = & 
 \left( \left. {\cal A}_{n+2}^{(0)} \right.^\ast {\cal A}_{n+2}^{(0)} \right) d\phi_{n+2},  
 \nonumber \\
d\sigma_{n+1}^{(1)} & = &  
 \left( 
 \left. {\cal A}_{n+1}^{(0)} \right.^\ast {\cal A}_{n+1}^{(1)} 
 + \left. {\cal A}_{n+1}^{(1)} \right.^\ast {\cal A}_{n+1}^{(0)} \right) d\phi_{n+1}, 
 \nonumber \\
d\sigma_n^{(2)} & = & 
 \left( 
 \left. {\cal A}_n^{(0)} \right.^\ast {\cal A}_n^{(2)} 
 + \left. {\cal A}_n^{(2)} \right.^\ast {\cal A}_n^{(0)}  
 + \left. {\cal A}_n^{(1)} \right.^\ast {\cal A}_n^{(1)} \right) d\phi_n, 
\eq
where ${\cal A}_n^{(l)}$ denotes an amplitude with $n$ external partons and $l$ loops.
$d\phi_n$ is the phase space measure for $n$ partons.
We would like to construct a numerical program for an arbitrary infrared safe observable $O$.
Several options for the cancellation of infrared divergences have been discussed
\cite{Kosower:2002su,Kosower:2003cz,Kosower:2003bh,Weinzierl:2003fx,Weinzierl:2003ra,Anastasiou:2003gr,Gehrmann-DeRidder:2003bm,Gehrmann-DeRidder:2004tv,Gehrmann-DeRidder:2005hi,Gehrmann-DeRidder:2005aw,Gehrmann-DeRidder:2005cm,Binoth:2004jv,Heinrich:2006sw,Kilgore:2004ty,Frixione:2004is,Catani:2007vq,Somogyi:2005xz,Somogyi:2006da,Somogyi:2006db}.
Among those, the subtraction method -- well-known from NLO computations
and sector decomposition 
are the most promising candidates.
Let us look again at the subtraction method.
To render the individual contributions finite, one adds and subtracts suitable
pieces:
\bq
\langle O \rangle_n^{NNLO} & = &
 \int 
               O_{n+2} \; d\sigma_{n+2}^{(0)} 
             - O_{n+1} \circ d\alpha^{(0,1)}_{n+1}
             - O_{n} \circ d\alpha^{(0,2)}_{n} 
 \nonumber \\
& &
 + \int 
                 O_{n+1} \; d\sigma_{n+1}^{(1)} 
               + O_{n+1} \circ d\alpha^{(0,1)}_{n+1}
               - O_{n} \circ d\alpha^{(1,1)}_{n}
 \nonumber \\
& & 
 + \int 
                 O_{n} \; d\sigma_n^{(2)} 
               + O_{n} \circ d\alpha^{(0,2)}_{n}
               + O_{n} \circ d\alpha^{(1,1)}_{n}.
 \nonumber
\eq
Here $d\alpha_{n+1}^{(0,1)}$ is a subtraction term for single unresolved configurations
of Born amplitudes.
This term is already known from NLO calculations.
The term $d\alpha_n^{(0,2)}$ is a subtraction term 
for double unresolved configurations.
Finally, $d\alpha_n^{(1,1)}$ is a subtraction term
for single unresolved configurations involving one-loop amplitudes.

To construct these terms the universal factorisation properties of 
QCD amplitudes in unresolved limits are essential.
QCD amplitudes factorise if they are decomposed into primitive
amplitudes.
Primitive amplitudes are defined by
a fixed cyclic ordering of the QCD partons,
a definite routing of the external fermion lines through the diagram
and the particle content circulating in the loop.
One-loop amplitudes factorise in single unresolved limits as
\cite{Bern:1994zx,Bern:1998sc,Kosower:1999xi,Kosower:1999rx,Bern:1999ry,Catani:2000pi,Kosower:2003cz}
\begin{eqnarray}
\label{oneloopfactformula}
A^{(1)}_{n}
  & = &
  \mbox{Sing}^{(0,1)} 
  \cdot A^{(1)}_{n-1} +
  \mbox{Sing}^{(1,1)} \cdot A^{(0)}_{n-1}.
\end{eqnarray}
Tree amplitudes factorise in the double unresolved limits as
\cite{Berends:1989zn,Gehrmann-DeRidder:1998gf,Campbell:1998hg,Catani:1998nv,Catani:1999ss,DelDuca:1999ha,Kosower:2002su}
\begin{eqnarray}
\label{factsing}
A^{(0)}_{n}
  & = &
  \mbox{Sing}^{(0,2)} \cdot A^{(0)}_{n-2}.
\end{eqnarray}
In addition, the pole structure of two-loop amplitudes is known \cite{Catani:1998bh,Sterman:2002qn,Mitov:2006xs}.
The subtraction terms have to interpolate between the various singular limits.
Spin-averaged antenna subtraction terms have been worked out in \cite{Gehrmann-DeRidder:2005cm,Weinzierl:2006ij}
and first results at NNLO based on the subtraction method are available for 
$e^+ e^- \rightarrow \mbox{2 jets}$ \cite{Weinzierl:2006ij,Weinzierl:2006yt}
and the thrust distribution \cite{Ridder:2007bj}.


\section{Summary}

Precision calculations for multi-parton processes will play an important role
for the physics program at the LHC.
I discussed various approaches how to overcome the bottle-necks we face technically: the length of expressions,
the calculation of loop integrals and the occurrence of infra-red divergences.
Computer algebra, a divide-and-conquer approach and new developments based on on-shell recursion
relations help us to keep the length of expressions manageable.
Sophisticated techniques related to Mellin-Barnes transformations, shuffle algebras or sector decomposition
allow us to compute multi-loop integrals.
Urgently needed are also automated computations of one-loop amplitudes. I discussed improvements of the
Passarino-Veltman algorithm and new ideas base on generalised unitarity.
As far as the cancellation of infrared divergences is concerned, there are systematic methods at NLO.
Several methods, which address the issue at NNLO have been proposed and first numerical results
for observables based on the use of these methods have emerged.

\subsection*{Acknowledgements}

The figures in this article have been produced with the help of the program axodraw \cite{Vermaseren:1994je}.

\bibliography{/home/stefanw/notes/biblio}

\providecommand{\href}[2]{#2}\begingroup\raggedright\begin{thebibliography}{10%
0}

\bibitem{Dittmaier:2007wz}
S.~Dittmaier, P.~Uwer, and S.~Weinzierl, {\it Nlo qcd corrections to t anti-t +
  jet production at hadron colliders},  {\em Phys. Rev. Lett.} {\bf 98} (2007)
  262002, [\href{http://xxx.lanl.gov/abs/hep-ph/0703120}{{\tt
  hep-ph/0703120}}].

\bibitem{Buttar:2006zd}
C.~Buttar {\em et~al.}, {\it Les houches physics at tev colliders 2005,
  standard model, qcd, ew, and higgs working group: Summary report},
  \href{http://xxx.lanl.gov/abs/hep-ph/0604120}{{\tt hep-ph/0604120}}.

\bibitem{Berends:1987me}
F.~A. Berends and W.~T. Giele, {\it Recursive calculations for processes with n
  gluons},  {\em Nucl. Phys.} {\bf B306} (1988) 759.

\bibitem{Berends:1989ie}
F.~A. Berends, W.~T. Giele, and H.~Kuijf, {\it On six jet production at hadron
  colliders},  {\em Phys. Lett.} {\bf B232} (1989) 266.

\bibitem{Berends:1990ax}
F.~A. Berends, H.~Kuijf, B.~Tausk, and W.~T. Giele, {\it On the production of a
  w and jets at hadron colliders},  {\em Nucl. Phys.} {\bf B357} (1991) 32--64.

\bibitem{Caravaglios:1995cd}
F.~Caravaglios and M.~Moretti, {\it An algorithm to compute born scattering
  amplitudes without feynman graphs},  {\em Phys. Lett.} {\bf B358} (1995)
  332--338, [\href{http://xxx.lanl.gov/abs/hep-ph/9507237}{{\tt
  hep-ph/9507237}}].

\bibitem{Caravaglios:1998yr}
F.~Caravaglios, M.~L. Mangano, M.~Moretti, and R.~Pittau, {\it A new approach
  to multi-jet calculations in hadron collisions},  {\em Nucl. Phys.} {\bf
  B539} (1999) 215--232, [\href{http://xxx.lanl.gov/abs/hep-ph/9807570}{{\tt
  hep-ph/9807570}}].

\bibitem{Draggiotis:1998gr}
P.~Draggiotis, R.~H.~P. Kleiss, and C.~G. Papadopoulos, {\it On the computation
  of multigluon amplitudes},  {\em Phys. Lett.} {\bf B439} (1998) 157--164,
  [\href{http://xxx.lanl.gov/abs/hep-ph/9807207}{{\tt hep-ph/9807207}}].

\bibitem{Draggiotis:2002hm}
P.~D. Draggiotis, R.~H.~P. Kleiss, and C.~G. Papadopoulos, {\it Multi-jet
  production in hadron collisions},  {\em Eur. Phys. J.} {\bf C24} (2002)
  447--458, [\href{http://xxx.lanl.gov/abs/hep-ph/0202201}{{\tt
  hep-ph/0202201}}].

\bibitem{Papadopoulos:2005ky}
C.~G. Papadopoulos and M.~Worek, {\it Multi-parton cross sections at hadron
  colliders},  {\em Eur. Phys. J.} {\bf C50} (2007) 843--856,
  [\href{http://xxx.lanl.gov/abs/hep-ph/0512150}{{\tt hep-ph/0512150}}].

\bibitem{Kleiss:1986gy}
R.~Kleiss, W.~J. Stirling, and S.~D. Ellis, {\it A new monte carlo treatment of
  multiparticle phase space at high-energies},  {\em Comput. Phys. Commun.}
  {\bf 40} (1986) 359.

\bibitem{vanHameren:2002tc}
A.~van Hameren and C.~G. Papadopoulos, {\it A hierarchical phase space
  generator for qcd antenna structures},  {\em Eur. Phys. J.} {\bf C25} (2002)
  563--574, [\href{http://xxx.lanl.gov/abs/hep-ph/0204055}{{\tt
  hep-ph/0204055}}].

\bibitem{Maltoni:2002qb}
F.~Maltoni and T.~Stelzer, {\it Madevent: Automatic event generation with
  madgraph},  {\em JHEP} {\bf 02} (2003) 027,
  [\href{http://xxx.lanl.gov/abs/hep-ph/0208156}{{\tt hep-ph/0208156}}].

\bibitem{Stelzer:1994ta}
T.~Stelzer and W.~F. Long, {\it Automatic generation of tree level helicity
  amplitudes},  {\em Comput. Phys. Commun.} {\bf 81} (1994) 357--371,
  [\href{http://xxx.lanl.gov/abs/hep-ph/9401258}{{\tt hep-ph/9401258}}].

\bibitem{Murayama:1992gi}
H.~Murayama, I.~Watanabe, and K.~Hagiwara, {\it Helas: Helicity amplitude
  subroutines for feynman diagram evaluations}, . KEK-91-11.

\bibitem{Krauss:2001iv}
F.~Krauss, R.~Kuhn, and G.~Soff, {\it Amegic++ 1.0: A matrix element generator
  in c++},  {\em JHEP} {\bf 02} (2002) 044,
  [\href{http://xxx.lanl.gov/abs/hep-ph/0109036}{{\tt hep-ph/0109036}}].

\bibitem{Kanaki:2000ey}
A.~Kanaki and C.~G. Papadopoulos, {\it Helac: A package to compute electroweak
  helicity amplitudes},  {\em Comput. Phys. Commun.} {\bf 132} (2000) 306--315,
  [\href{http://xxx.lanl.gov/abs/hep-ph/0002082}{{\tt hep-ph/0002082}}].

\bibitem{Papadopoulos:2000tt}
C.~G. Papadopoulos, {\it Phegas: A phase space generator for automatic
  cross-section computation},  {\em Comput. Phys. Commun.} {\bf 137} (2001)
  247--254, [\href{http://xxx.lanl.gov/abs/hep-ph/0007335}{{\tt
  hep-ph/0007335}}].

\bibitem{Pukhov:1999gg}
A.~Pukhov {\em et~al.}, {\it Comphep: A package for evaluation of feynman
  diagrams and integration over multi-particle phase space. user's manual for
  version 33},  \href{http://xxx.lanl.gov/abs/hep-ph/9908288}{{\tt
  hep-ph/9908288}}.

\bibitem{Yuasa:1999rg}
F.~Yuasa {\em et~al.}, {\it Automatic computation of cross sections in hep:
  Status of grace system},  {\em Prog. Theor. Phys. Suppl.} {\bf 138} (2000)
  18--23, [\href{http://xxx.lanl.gov/abs/hep-ph/0007053}{{\tt
  hep-ph/0007053}}].

\bibitem{Mangano:2002ea}
M.~L. Mangano, M.~Moretti, F.~Piccinini, R.~Pittau, and A.~D. Polosa, {\it
  Alpgen, a generator for hard multiparton processes in hadronic collisions},
  {\em JHEP} {\bf 07} (2003) 001,
  [\href{http://xxx.lanl.gov/abs/hep-ph/0206293}{{\tt hep-ph/0206293}}].

\bibitem{Kilgore:1996sq}
W.~B. Kilgore and W.~T. Giele, {\it Next-to-leading order gluonic three jet
  production at hadron colliders},  {\em Phys. Rev.} {\bf D55} (1997)
  7183--7190, [\href{http://xxx.lanl.gov/abs/hep-ph/9610433}{{\tt
  hep-ph/9610433}}].

\bibitem{Nagy:2001fj}
Z.~Nagy, {\it Three-jet cross sections in hadron hadron collisions at
  next-to-leading order},  {\em Phys. Rev. Lett.} {\bf 88} (2002) 122003,
  [\href{http://xxx.lanl.gov/abs/hep-ph/0110315}{{\tt hep-ph/0110315}}].

\bibitem{Nagy:2003tz}
Z.~Nagy, {\it Next-to-leading order calculation of three-jet observables in
  hadron hadron collision},  {\em Phys. Rev.} {\bf D68} (2003) 094002,
  [\href{http://xxx.lanl.gov/abs/hep-ph/0307268}{{\tt hep-ph/0307268}}].

\bibitem{Campbell:2000bg}
J.~M. Campbell and R.~K. Ellis, {\it Radiative corrections to z b anti-b
  production},  {\em Phys. Rev.} {\bf D62} (2000) 114012,
  [\href{http://xxx.lanl.gov/abs/hep-ph/0006304}{{\tt hep-ph/0006304}}].

\bibitem{Campbell:2002tg}
J.~Campbell and R.~K. Ellis, {\it Next-to-leading order corrections to w + 2jet
  and z + 2jet production at hadron colliders},  {\em Phys. Rev.} {\bf D65}
  (2002) 113007, [\href{http://xxx.lanl.gov/abs/hep-ph/0202176}{{\tt
  hep-ph/0202176}}].

\bibitem{Campbell:2003hd}
J.~Campbell, R.~K. Ellis, and D.~L. Rainwater, {\it Next-to-leading order qcd
  predictions for w + 2jet and z + 2jet production at the cern lhc},  {\em
  Phys. Rev.} {\bf D68} (2003) 094021,
  [\href{http://xxx.lanl.gov/abs/hep-ph/0308195}{{\tt hep-ph/0308195}}].

\bibitem{Campbell:2005zv}
J.~Campbell, R.~K. Ellis, F.~Maltoni, and S.~Willenbrock, {\it Production of a
  z boson and two jets with one heavy-quark tag},  {\em Phys. Rev.} {\bf D73}
  (2006) 054007, [\href{http://xxx.lanl.gov/abs/hep-ph/0510362}{{\tt
  hep-ph/0510362}}].

\bibitem{Campbell:2006cu}
J.~Campbell, R.~K. Ellis, F.~Maltoni, and S.~Willenbrock, {\it Production of a
  w boson and two jets with one b-quark tag},  {\em Phys. Rev.} {\bf D75}
  (2007) 054015, [\href{http://xxx.lanl.gov/abs/hep-ph/0611348}{{\tt
  hep-ph/0611348}}].

\bibitem{Beenakker:2002nc}
W.~Beenakker {\em et~al.}, {\it Nlo qcd corrections to t anti-t h production in
  hadron collisions. ((u))},  {\em Nucl. Phys.} {\bf B653} (2003) 151--203,
  [\href{http://xxx.lanl.gov/abs/hep-ph/0211352}{{\tt hep-ph/0211352}}].

\bibitem{Dawson:2003zu}
S.~Dawson, C.~Jackson, L.~H. Orr, L.~Reina, and D.~Wackeroth, {\it Associated
  higgs production with top quarks at the large hadron collider: Nlo qcd
  corrections},  {\em Phys. Rev.} {\bf D68} (2003) 034022,
  [\href{http://xxx.lanl.gov/abs/hep-ph/0305087}{{\tt hep-ph/0305087}}].

\bibitem{DelDuca:2001eu}
V.~Del~Duca, W.~Kilgore, C.~Oleari, C.~Schmidt, and D.~Zeppenfeld, {\it H + 2
  jets via gluon fusion},  {\em Phys. Rev. Lett.} {\bf 87} (2001) 122001,
  [\href{http://xxx.lanl.gov/abs/hep-ph/0105129}{{\tt hep-ph/0105129}}].

\bibitem{DelDuca:2001fn}
V.~Del~Duca, W.~Kilgore, C.~Oleari, C.~Schmidt, and D.~Zeppenfeld, {\it
  Gluon-fusion contributions to h + 2 jet production},  {\em Nucl. Phys.} {\bf
  B616} (2001) 367--399, [\href{http://xxx.lanl.gov/abs/hep-ph/0108030}{{\tt
  hep-ph/0108030}}].

\bibitem{Campbell:2006xx}
J.~M. Campbell, R.~Keith~Ellis, and G.~Zanderighi, {\it Next-to-leading order
  higgs + 2 jet production via gluon fusion},  {\em JHEP} {\bf 10} (2006) 028,
  [\href{http://xxx.lanl.gov/abs/hep-ph/0608194}{{\tt hep-ph/0608194}}].

\bibitem{Lazopoulos:2007ix}
A.~Lazopoulos, K.~Melnikov, and F.~Petriello, {\it Qcd corrections to tri-boson
  production},  {\em Phys. Rev.} {\bf D76} (2007) 014001,
  [\href{http://xxx.lanl.gov/abs/hep-ph/0703273}{{\tt hep-ph/0703273}}].

\bibitem{Jager:2006zc}
B.~Jager, C.~Oleari, and D.~Zeppenfeld, {\it Next-to-leading order qcd
  corrections to w+ w- production via vector-boson fusion},  {\em JHEP} {\bf
  07} (2006) 015, [\href{http://xxx.lanl.gov/abs/hep-ph/0603177}{{\tt
  hep-ph/0603177}}].

\bibitem{Jager:2006cp}
B.~Jager, C.~Oleari, and D.~Zeppenfeld, {\it Next-to-leading order qcd
  corrections to z boson pair production via vector-boson fusion},  {\em Phys.
  Rev.} {\bf D73} (2006) 113006,
  [\href{http://xxx.lanl.gov/abs/hep-ph/0604200}{{\tt hep-ph/0604200}}].

\bibitem{Bozzi:2007ur}
G.~Bozzi, B.~Jager, C.~Oleari, and D.~Zeppenfeld, {\it Next-to-leading order
  qcd corrections to w+z and w-z production via vector-boson fusion},  {\em
  Phys. Rev.} {\bf D75} (2007) 073004,
  [\href{http://xxx.lanl.gov/abs/hep-ph/0701105}{{\tt hep-ph/0701105}}].

\bibitem{Denner:2005es}
A.~Denner, S.~Dittmaier, M.~Roth, and L.~H. Wieders, {\it Complete electroweak
  o(alpha) corrections to charged- current e+ e- --> 4fermion processes},  {\em
  Phys. Lett.} {\bf B612} (2005) 223--232,
  [\href{http://xxx.lanl.gov/abs/hep-ph/0502063}{{\tt hep-ph/0502063}}].

\bibitem{Denner:2005fg}
A.~Denner, S.~Dittmaier, M.~Roth, and L.~H. Wieders, {\it Electroweak
  corrections to charged-current e+ e- --> 4 fermion processes: Technical
  details and further results},  {\em Nucl. Phys.} {\bf B724} (2005) 247--294,
  [\href{http://xxx.lanl.gov/abs/hep-ph/0505042}{{\tt hep-ph/0505042}}].

\bibitem{Dixon:1997th}
L.~Dixon and A.~Signer, {\it Complete o (alpha-s**3) results for e+ e- $\to$
  (gamma, z) $\to$ four jets},  {\em Phys. Rev.} {\bf D56} (1997) 4031--4038,
  [\href{http://xxx.lanl.gov/abs/hep-ph/9706285}{{\tt hep-ph/9706285}}].

\bibitem{Nagy:1997yn}
Z.~Nagy and Z.~Trocsanyi, {\it Next-to-leading order calculation of four jet
  shape variables},  {\em Phys. Rev. Lett.} {\bf 79} (1997) 3604--3607,
  [\href{http://xxx.lanl.gov/abs/hep-ph/9707309}{{\tt hep-ph/9707309}}].

\bibitem{Campbell:1998nn}
J.~M. Campbell, M.~A. Cullen, and E.~W.~N. Glover, {\it Four jet event shapes
  in electron - positron annihilation},  {\em Eur. Phys. J.} {\bf C9} (1999)
  245, [\href{http://xxx.lanl.gov/abs/hep-ph/9809429}{{\tt hep-ph/9809429}}].

\bibitem{Weinzierl:1999yf}
S.~Weinzierl and D.~A. Kosower, {\it {QCD} corrections to four-jet production
  and three-jet structure in e+ e- annihilation},  {\em Phys. Rev.} {\bf D60}
  (1999) 054028, [\href{http://xxx.lanl.gov/abs/hep-ph/9901277}{{\tt
  hep-ph/9901277}}].

\bibitem{Kublbeck:1990xc}
J.~K{\"u}blbeck, M.~Bohm, and A.~Denner, {\it Feyn arts: Computer algebraic
  generation of feynman graphs and amplitudes},  {\em Comput. Phys. Commun.}
  {\bf 60} (1990) 165--180.

\bibitem{Hahn:2000kx}
T.~Hahn, {\it Generating feynman diagrams and amplitudes with feynarts 3},
  {\em Comput. Phys. Commun.} {\bf 140} (2001) 418--431,
  [\href{http://xxx.lanl.gov/abs/hep-ph/0012260}{{\tt hep-ph/0012260}}].

\bibitem{Hahn:1998yk}
T.~Hahn and M.~Perez-Victoria, {\it Automatized one-loop calculations in four
  and d dimensions},  {\em Comput. Phys. Commun.} {\bf 118} (1999) 153--165,
  [\href{http://xxx.lanl.gov/abs/hep-ph/9807565}{{\tt hep-ph/9807565}}].

\bibitem{vanOldenborgh:1990yc}
G.~J. van Oldenborgh, {\it Ff: A package to evaluate one loop feynman
  diagrams},  {\em Comput. Phys. Commun.} {\bf 66} (1991) 1--15.

\bibitem{Belanger:2003sd}
G.~Belanger {\em et~al.}, {\it Automatic calculations in high energy physics
  and grace at one-loop},  {\em Phys. Rept.} {\bf 430} (2006) 117--209,
  [\href{http://xxx.lanl.gov/abs/hep-ph/0308080}{{\tt hep-ph/0308080}}].

\bibitem{Anastasiou:2003yy}
C.~Anastasiou, L.~J. Dixon, K.~Melnikov, and F.~Petriello, {\it Dilepton
  rapidity distribution in the drell-yan process at nnlo in qcd},  {\em Phys.
  Rev. Lett.} {\bf 91} (2003) 182002,
  [\href{http://xxx.lanl.gov/abs/hep-ph/0306192}{{\tt hep-ph/0306192}}].

\bibitem{Anastasiou:2003ds}
C.~Anastasiou, L.~Dixon, K.~Melnikov, and F.~Petriello, {\it High-precision qcd
  at hadron colliders: Electroweak gauge boson rapidity distributions at nnlo},
   {\em Phys. Rev.} {\bf D69} (2004) 094008,
  [\href{http://xxx.lanl.gov/abs/hep-ph/0312266}{{\tt hep-ph/0312266}}].

\bibitem{Anastasiou:2005qj}
C.~Anastasiou, K.~Melnikov, and F.~Petriello, {\it Fully differential higgs
  boson production and the di-photon signal through next-to-next-to-leading
  order},  {\em Nucl. Phys.} {\bf B724} (2005) 197--246,
  [\href{http://xxx.lanl.gov/abs/hep-ph/0501130}{{\tt hep-ph/0501130}}].

\bibitem{Anastasiou:2004xq}
C.~Anastasiou, K.~Melnikov, and F.~Petriello, {\it Higgs boson production at
  hadron colliders: Differential cross sections through next-to-next-to-leading
  order},  {\em Phys. Rev. Lett.} {\bf 93} (2004) 262002,
  [\href{http://xxx.lanl.gov/abs/hep-ph/0409088}{{\tt hep-ph/0409088}}].

\bibitem{Anastasiou:2002yz}
C.~Anastasiou and K.~Melnikov, {\it Higgs boson production at hadron colliders
  in nnlo qcd},  {\em Nucl. Phys.} {\bf B646} (2002) 220--256,
  [\href{http://xxx.lanl.gov/abs/hep-ph/0207004}{{\tt hep-ph/0207004}}].

\bibitem{Harlander:2002wh}
R.~V. Harlander and W.~B. Kilgore, {\it Next-to-next-to-leading order higgs
  production at hadron colliders},  {\em Phys. Rev. Lett.} {\bf 88} (2002)
  201801, [\href{http://xxx.lanl.gov/abs/hep-ph/0201206}{{\tt
  hep-ph/0201206}}].

\bibitem{Ravindran:2003um}
V.~Ravindran, J.~Smith, and W.~L. van Neerven, {\it Nnlo corrections to the
  total cross section for higgs boson production in hadron hadron collisions},
  {\em Nucl. Phys.} {\bf B665} (2003) 325--366,
  [\href{http://xxx.lanl.gov/abs/hep-ph/0302135}{{\tt hep-ph/0302135}}].

\bibitem{Harlander:2001is}
R.~V. Harlander and W.~B. Kilgore, {\it Soft and virtual corrections to p p -->
  h + x at nnlo},  {\em Phys. Rev.} {\bf D64} (2001) 013015,
  [\href{http://xxx.lanl.gov/abs/hep-ph/0102241}{{\tt hep-ph/0102241}}].

\bibitem{Catani:2001ic}
S.~Catani, D.~de~Florian, and M.~Grazzini, {\it Higgs production in hadron
  collisions: Soft and virtual qcd corrections at nnlo},  {\em JHEP} {\bf 05}
  (2001) 025, [\href{http://xxx.lanl.gov/abs/hep-ph/0102227}{{\tt
  hep-ph/0102227}}].

\bibitem{Anastasiou:2004qd}
C.~Anastasiou, K.~Melnikov, and F.~Petriello, {\it Real radiation at nnlo: e+
  e- --> 2jets through o(alpha(s)**2)},  {\em Phys. Rev. Lett.} {\bf 93} (2004)
  032002, [\href{http://xxx.lanl.gov/abs/hep-ph/0402280}{{\tt
  hep-ph/0402280}}].

\bibitem{Weinzierl:2006ij}
S.~Weinzierl, {\it Nnlo corrections to 2-jet observables in electron positron
  annihilation},  {\em Phys. Rev.} {\bf D74} (2006) 014020,
  [\href{http://xxx.lanl.gov/abs/hep-ph/0606008}{{\tt hep-ph/0606008}}].

\bibitem{Weinzierl:2006yt}
S.~Weinzierl, {\it The forward-backward asymmetry at nnlo revisited},  {\em
  Phys. Lett.} {\bf B644} (2007) 331--335,
  [\href{http://xxx.lanl.gov/abs/hep-ph/0609021}{{\tt hep-ph/0609021}}].

\bibitem{Ridder:2007bj}
A.~G.-D. Ridder, T.~Gehrmann, E.~W.~N. Glover, and G.~Heinrich, {\it
  Second-order qcd corrections to the thrust distribution},
  \href{http://xxx.lanl.gov/abs/arXiv:0707.1285 [hep-ph]}{{\tt arXiv:0707.1285
  [hep-ph]}}.

\bibitem{Vermaseren:2000nd}
J.~A.~M. Vermaseren, {\it New features of form},
  \href{http://xxx.lanl.gov/abs/math-ph/0010025}{{\tt math-ph/0010025}}.

\bibitem{Bauer:2000cp}
C.~Bauer, A.~Frink, and R.~Kreckel, {\it Introduction to the ginac framework
  for symbolic computation within the c++ programming language},  {\em J.
  Symbolic Computation} {\bf 33} (2002) 1,
  [\href{http://xxx.lanl.gov/abs/cs.sc/0004015}{{\tt cs.sc/0004015}}].

\bibitem{Vollinga:2005pk}
J.~Vollinga, {\it Ginac: Symbolic computation with c++},  {\em Nucl. Instrum.
  Meth.} {\bf A559} (2006) 282--284,
  [\href{http://xxx.lanl.gov/abs/hep-ph/0510057}{{\tt hep-ph/0510057}}].

\bibitem{Berends:1981rb}
F.~A. Berends, R.~Kleiss, P.~De~Causmaecker, R.~Gastmans, and T.~T. Wu, {\it
  Single bremsstrahlung processes in gauge theories},  {\em Phys. Lett.} {\bf
  B103} (1981) 124.

\bibitem{DeCausmaecker:1982bg}
P.~De~Causmaecker, R.~Gastmans, W.~Troost, and T.~T. Wu, {\it Multiple
  bremsstrahlung in gauge theories at high-energies. 1. general formalism for
  quantum electrodynamics},  {\em Nucl. Phys.} {\bf B206} (1982) 53.

\bibitem{Gunion:1985vc}
J.~F. Gunion and Z.~Kunszt, {\it Improved analytic techniques for tree graph
  calculations and the g g q anti-q lepton anti-lepton subprocess},  {\em Phys.
  Lett.} {\bf B161} (1985) 333.

\bibitem{Kleiss:1986qc}
R.~Kleiss and W.~J. Stirling, {\it Cross-sections for the production of an
  arbitrary number of photons in electron - positron annihilation},  {\em Phys.
  Lett.} {\bf B179} (1986) 159.

\bibitem{Xu:1987xb}
Z.~Xu, D.-H. Zhang, and L.~Chang, {\it Helicity amplitudes for multiple
  bremsstrahlung in massless nonabelian gauge theories},  {\em Nucl. Phys.}
  {\bf B291} (1987) 392.

\bibitem{Gastmans:1990xh}
R.~Gastmans and T.~T. Wu, {\it The ubiquitous photon: Helicity method for qed
  and qcd}, . Oxford, UK: Clarendon (1990) 648 p. (International series of
  monographs on physics, 80).

\bibitem{Schwinn:2005pi}
C.~Schwinn and S.~Weinzierl, {\it Scalar diagrammatic rules for born amplitudes
  in qcd},  {\em JHEP} {\bf 05} (2005) 006,
  [\href{http://xxx.lanl.gov/abs/hep-th/0503015}{{\tt hep-th/0503015}}].

\bibitem{Rodrigo:2005eu}
G.~Rodrigo, {\it Multigluonic scattering amplitudes of heavy quarks},  {\em
  JHEP} {\bf 09} (2005) 079,
  [\href{http://xxx.lanl.gov/abs/hep-ph/0508138}{{\tt hep-ph/0508138}}].

\bibitem{Cvitanovic:1980bu}
P.~Cvitanovic, P.~G. Lauwers, and P.~N. Scharbach, {\it Gauge invariance
  structure of quantum chromodynamics},  {\em Nucl. Phys.} {\bf B186} (1981)
  165.

\bibitem{Berends:1987cv}
F.~A. Berends and W.~Giele, {\it The six gluon process as an example of
  weyl-van der waerden spinor calculus},  {\em Nucl. Phys.} {\bf B294} (1987)
  700.

\bibitem{Mangano:1987xk}
M.~L. Mangano, S.~J. Parke, and Z.~Xu, {\it Duality and multi - gluon
  scattering},  {\em Nucl. Phys.} {\bf B298} (1988) 653.

\bibitem{Kosower:1987ic}
D.~Kosower, B.-H. Lee, and V.~P. Nair, {\it Multi gluon scattering: A string
  based calculation},  {\em Phys. Lett.} {\bf B201} (1988) 85.

\bibitem{Bern:1990ux}
Z.~Bern and D.~A. Kosower, {\it Color decomposition of one loop amplitudes in
  gauge theories},  {\em Nucl. Phys.} {\bf B362} (1991) 389--448.

\bibitem{DelDuca:1999rs}
V.~Del~Duca, L.~J. Dixon, and F.~Maltoni, {\it New color decompositions for
  gauge amplitudes at tree and loop level},  {\em Nucl. Phys.} {\bf B571}
  (2000) 51--70, [\href{http://xxx.lanl.gov/abs/hep-ph/9910563}{{\tt
  hep-ph/9910563}}].

\bibitem{Maltoni:2002mq}
F.~Maltoni, K.~Paul, T.~Stelzer, and S.~Willenbrock, {\it Color-flow
  decomposition of qcd amplitudes},  {\em Phys. Rev.} {\bf D67} (2003) 014026,
  [\href{http://xxx.lanl.gov/abs/hep-ph/0209271}{{\tt hep-ph/0209271}}].

\bibitem{Weinzierl:2005dd}
S.~Weinzierl, {\it Automated computation of spin- and colour-correlated born
  matrix elements},  {\em Eur. Phys. J.} {\bf C45} (2006) 745--757,
  [\href{http://xxx.lanl.gov/abs/hep-ph/0510157}{{\tt hep-ph/0510157}}].

\bibitem{Kosower:1989xy}
D.~A. Kosower, {\it Light cone recurrence relations for qcd amplitudes},  {\em
  Nucl. Phys.} {\bf B335} (1990) 23.

\bibitem{Parke:1986gb}
S.~J. Parke and T.~R. Taylor, {\it An amplitude for n gluon scattering},  {\em
  Phys. Rev. Lett.} {\bf 56} (1986) 2459.

\bibitem{Grisaru:1976vm}
M.~T. Grisaru, H.~N. Pendleton, and P.~van Nieuwenhuizen, {\it Supergravity and
  the s matrix},  {\em Phys. Rev.} {\bf D15} (1977) 996.

\bibitem{Grisaru:1977px}
M.~T. Grisaru and H.~N. Pendleton, {\it Some properties of scattering
  amplitudes in supersymmetric theories},  {\em Nucl. Phys.} {\bf B124} (1977)
  81.

\bibitem{Parke:1985pn}
S.~J. Parke and T.~R. Taylor, {\it Perturbative qcd utilizing extended
  supersymmetry},  {\em Phys. Lett.} {\bf B157} (1985) 81.

\bibitem{Reuter:2002gn}
J.~Reuter, {\it Supersymmetry of scattering amplitudes and green functions in
  perturbation theory},  \href{http://xxx.lanl.gov/abs/hep-th/0212154}{{\tt
  hep-th/0212154}}.

\bibitem{Schwinn:2006ca}
C.~Schwinn and S.~Weinzierl, {\it Susy ward identities for multi-gluon helicity
  amplitudes with massive quarks},  {\em JHEP} {\bf 03} (2006) 030,
  [\href{http://xxx.lanl.gov/abs/hep-th/0602012}{{\tt hep-th/0602012}}].

\bibitem{Cachazo:2004kj}
F.~Cachazo, P.~Svrcek, and E.~Witten, {\it Mhv vertices and tree amplitudes in
  gauge theory},  {\em JHEP} {\bf 09} (2004) 006,
  [\href{http://xxx.lanl.gov/abs/hep-th/0403047}{{\tt hep-th/0403047}}].

\bibitem{Kosower:2004yz}
D.~A. Kosower, {\it Next-to-maximal helicity violating amplitudes in gauge
  theory},  {\em Phys. Rev.} {\bf D71} (2005) 045007,
  [\href{http://xxx.lanl.gov/abs/hep-th/0406175}{{\tt hep-th/0406175}}].

\bibitem{Britto:2004ap}
R.~Britto, F.~Cachazo, and B.~Feng, {\it New recursion relations for tree
  amplitudes of gluons},  {\em Nucl. Phys.} {\bf B715} (2005) 499--522,
  [\href{http://xxx.lanl.gov/abs/hep-th/0412308}{{\tt hep-th/0412308}}].

\bibitem{Dinsdale:2006sq}
M.~Dinsdale, M.~Ternick, and S.~Weinzierl, {\it A comparison of efficient
  methods for the computation of born gluon amplitudes},  {\em JHEP} {\bf 03}
  (2006) 056, [\href{http://xxx.lanl.gov/abs/hep-ph/0602204}{{\tt
  hep-ph/0602204}}].

\bibitem{Duhr:2006iq}
C.~Duhr, S.~Hoche, and F.~Maltoni, {\it Color-dressed recursive relations for
  multi-parton amplitudes},  {\em JHEP} {\bf 08} (2006) 062,
  [\href{http://xxx.lanl.gov/abs/hep-ph/0607057}{{\tt hep-ph/0607057}}].

\bibitem{Britto:2005fq}
R.~Britto, F.~Cachazo, B.~Feng, and E.~Witten, {\it Direct proof of tree-level
  recursion relation in yang-mills theory},  {\em Phys. Rev. Lett.} {\bf 94}
  (2005) 181602, [\href{http://xxx.lanl.gov/abs/hep-th/0501052}{{\tt
  hep-th/0501052}}].

\bibitem{Badger:2005zh}
S.~D. Badger, E.~W.~N. Glover, V.~V. Khoze, and P.~Svrcek, {\it Recursion
  relations for gauge theory amplitudes with massive particles},  {\em JHEP}
  {\bf 07} (2005) 025, [\href{http://xxx.lanl.gov/abs/hep-th/0504159}{{\tt
  hep-th/0504159}}].

\bibitem{Risager:2005vk}
K.~Risager, {\it A direct proof of the csw rules},  {\em JHEP} {\bf 12} (2005)
  003, [\href{http://xxx.lanl.gov/abs/hep-th/0508206}{{\tt hep-th/0508206}}].

\bibitem{Draggiotis:2005wq}
P.~D. Draggiotis, R.~H.~P. Kleiss, A.~Lazopoulos, and C.~G. Papadopoulos, {\it
  Diagrammatic proof of the bcfw recursion relation for gluon amplitudes in
  qcd},  {\em Eur. Phys. J.} {\bf C46} (2006) 741,
  [\href{http://xxx.lanl.gov/abs/hep-ph/0511288}{{\tt hep-ph/0511288}}].

\bibitem{Vaman:2005dt}
D.~Vaman and Y.-P. Yao, {\it Qcd recursion relations from the largest time
  equation},  {\em JHEP} {\bf 04} (2006) 030,
  [\href{http://xxx.lanl.gov/abs/hep-th/0512031}{{\tt hep-th/0512031}}].

\bibitem{Schwinn:2007ee}
C.~Schwinn and S.~Weinzierl, {\it On-shell recursion relations for all born qcd
  amplitudes},  {\em JHEP} {\bf 04} (2007) 072,
  [\href{http://xxx.lanl.gov/abs/hep-ph/0703021}{{\tt hep-ph/0703021}}].

\bibitem{Melrose:1965kb}
D.~B. Melrose, {\it Reduction of feynman diagrams},  {\em Nuovo Cim.} {\bf 40}
  (1965) 181--213.

\bibitem{vanNeerven:1984vr}
W.~L. van Neerven and J.~A.~M. Vermaseren, {\it Large loop integrals},  {\em
  Phys. Lett.} {\bf B137} (1984) 241.

\bibitem{Bern:1994kr}
Z.~Bern, L.~J. Dixon, and D.~A. Kosower, {\it Dimensionally regulated pentagon
  integrals},  {\em Nucl. Phys.} {\bf B412} (1994) 751--816,
  [\href{http://xxx.lanl.gov/abs/hep-ph/9306240}{{\tt hep-ph/9306240}}].

\bibitem{Binoth:1999sp}
T.~Binoth, J.~P. Guillet, and G.~Heinrich, {\it Reduction formalism for
  dimensionally regulated one-loop n- point integrals},  {\em Nucl. Phys.} {\bf
  B572} (2000) 361--386, [\href{http://xxx.lanl.gov/abs/hep-ph/9911342}{{\tt
  hep-ph/9911342}}].

\bibitem{Fleischer:1999hq}
J.~Fleischer, F.~Jegerlehner, and O.~V. Tarasov, {\it Algebraic reduction of
  one-loop feynman graph amplitudes},  {\em Nucl. Phys.} {\bf B566} (2000)
  423--440, [\href{http://xxx.lanl.gov/abs/hep-ph/9907327}{{\tt
  hep-ph/9907327}}].

\bibitem{Denner:2002ii}
A.~Denner and S.~Dittmaier, {\it Reduction of one-loop tensor 5-point
  integrals},  {\em Nucl. Phys.} {\bf B658} (2003) 175--202,
  [\href{http://xxx.lanl.gov/abs/hep-ph/0212259}{{\tt hep-ph/0212259}}].

\bibitem{Duplancic:2003tv}
G.~Duplancic and B.~Nizic, {\it Reduction method for dimensionally regulated
  one-loop n- point feynman integrals},  {\em Eur. Phys. J.} {\bf C35} (2004)
  105--118, [\href{http://xxx.lanl.gov/abs/hep-ph/0303184}{{\tt
  hep-ph/0303184}}].

\bibitem{Giele:2004iy}
W.~T. Giele and E.~W.~N. Glover, {\it A calculational formalism for one-loop
  integrals},  {\em JHEP} {\bf 04} (2004) 029,
  [\href{http://xxx.lanl.gov/abs/hep-ph/0402152}{{\tt hep-ph/0402152}}].

\bibitem{Passarino:1979jh}
G.~Passarino and M.~J.~G. Veltman, {\it One loop corrections for e+ e-
  annihilation into mu+ mu- in the weinberg model},  {\em Nucl. Phys.} {\bf
  B160} (1979) 151.

\bibitem{Denner:2005nn}
A.~Denner and S.~Dittmaier, {\it Reduction schemes for one-loop tensor
  integrals},  {\em Nucl. Phys.} {\bf B734} (2006) 62--115,
  [\href{http://xxx.lanl.gov/abs/hep-ph/0509141}{{\tt hep-ph/0509141}}].

\bibitem{Dittmaier:2003bc}
S.~Dittmaier, {\it Separation of soft and collinear singularities from one-
  loop n-point integrals},  {\em Nucl. Phys.} {\bf B675} (2003) 447--466,
  [\href{http://xxx.lanl.gov/abs/hep-ph/0308246}{{\tt hep-ph/0308246}}].

\bibitem{Stuart:1990de}
R.~G. Stuart and A.~Gongora, {\it Algebraic reduction of one loop feynman
  diagrams to scalar integrals. 2},  {\em Comput. Phys. Commun.} {\bf 56}
  (1990) 337--350.

\bibitem{Stuart:1988tt}
R.~G. Stuart, {\it Algebraic reduction of one loop feynman diagrams to scalar
  integrals},  {\em Comput. Phys. Commun.} {\bf 48} (1988) 367--389.

\bibitem{Davydychev:1991va}
A.~I. Davydychev, {\it A simple formula for reducing feynman diagrams to scalar
  integrals},  {\em Phys. Lett.} {\bf B263} (1991) 107--111.

\bibitem{Bern:1992em}
Z.~Bern, L.~J. Dixon, and D.~A. Kosower, {\it Dimensionally regulated one loop
  integrals},  {\em Phys. Lett.} {\bf B302} (1993) 299--308,
  [\href{http://xxx.lanl.gov/abs/hep-ph/9212308}{{\tt hep-ph/9212308}}].

\bibitem{Bern:1993kr}
Z.~Bern, L.~J. Dixon, and D.~A. Kosower, {\it Dimensionally regulated pentagon
  integrals},  {\em Nucl. Phys.} {\bf B412} (1994) 751--816,
  [\href{http://xxx.lanl.gov/abs/hep-ph/9306240}{{\tt hep-ph/9306240}}].

\bibitem{Campbell:1997zw}
J.~M. Campbell, E.~W.~N. Glover, and D.~J. Miller, {\it One-loop tensor
  integrals in dimensional regularisation},  {\em Nucl. Phys.} {\bf B498}
  (1997) 397--442, [\href{http://xxx.lanl.gov/abs/hep-ph/9612413}{{\tt
  hep-ph/9612413}}].

\bibitem{Giele:2004ub}
W.~Giele, E.~W.~N. Glover, and G.~Zanderighi, {\it Numerical evaluation of
  one-loop diagrams near exceptional momentum configurations},  {\em Nucl.
  Phys. Proc. Suppl.} {\bf 135} (2004) 275--279,
  [\href{http://xxx.lanl.gov/abs/hep-ph/0407016}{{\tt hep-ph/0407016}}].

\bibitem{Ellis:2005zh}
R.~K. Ellis, W.~T. Giele, and G.~Zanderighi, {\it Semi-numerical evaluation of
  one-loop corrections},  {\em Phys. Rev.} {\bf D73} (2006) 014027,
  [\href{http://xxx.lanl.gov/abs/hep-ph/0508308}{{\tt hep-ph/0508308}}].

\bibitem{Ferroglia:2002mz}
A.~Ferroglia, M.~Passera, G.~Passarino, and S.~Uccirati, {\it All-purpose
  numerical evaluation of one-loop multi-leg feynman diagrams},  {\em Nucl.
  Phys.} {\bf B650} (2003) 162--228,
  [\href{http://xxx.lanl.gov/abs/hep-ph/0209219}{{\tt hep-ph/0209219}}].

\bibitem{Kramer:2002cd}
M.~Kr{\"a}mer and D.~E. Soper, {\it Next-to-leading order numerical
  calculations in coulomb gauge},  {\em Phys. Rev.} {\bf D66} (2002) 054017,
  [\href{http://xxx.lanl.gov/abs/hep-ph/0204113}{{\tt hep-ph/0204113}}].

\bibitem{Nagy:2003qn}
Z.~Nagy and D.~E. Soper, {\it General subtraction method for numerical
  calculation of one-loop qcd matrix elements},  {\em JHEP} {\bf 09} (2003)
  055, [\href{http://xxx.lanl.gov/abs/hep-ph/0308127}{{\tt hep-ph/0308127}}].

\bibitem{Binoth:2002xh}
T.~Binoth, G.~Heinrich, and N.~Kauer, {\it A numerical evaluation of the scalar
  hexagon integral in the physical region},  {\em Nucl. Phys.} {\bf B654}
  (2003) 277--300, [\href{http://xxx.lanl.gov/abs/hep-ph/0210023}{{\tt
  hep-ph/0210023}}].

\bibitem{Binoth:2005ff}
T.~Binoth, J.~P. Guillet, G.~Heinrich, E.~Pilon, and C.~Schubert, {\it An
  algebraic / numerical formalism for one-loop multi-leg amplitudes},  {\em
  JHEP} {\bf 10} (2005) 015,
  [\href{http://xxx.lanl.gov/abs/hep-ph/0504267}{{\tt hep-ph/0504267}}].

\bibitem{Anastasiou:2007qb}
C.~Anastasiou, S.~Beerli, and A.~Daleo, {\it Evaluating multi-loop feynman
  diagrams with infrared and threshold singularities numerically},  {\em JHEP}
  {\bf 05} (2007) 071, [\href{http://xxx.lanl.gov/abs/hep-ph/0703282}{{\tt
  hep-ph/0703282}}].

\bibitem{Pittau:1997ez}
R.~Pittau, {\it A simple method for multi-leg loop calculations},  {\em Comput.
  Phys. Commun.} {\bf 104} (1997) 23--36,
  [\href{http://xxx.lanl.gov/abs/hep-ph/9607309}{{\tt hep-ph/9607309}}].

\bibitem{Pittau:1998mv}
R.~Pittau, {\it A simple method for multi-leg loop calculations. ii: A general
  algorithm},  {\em Comput. Phys. Commun.} {\bf 111} (1998) 48--52,
  [\href{http://xxx.lanl.gov/abs/hep-ph/9712418}{{\tt hep-ph/9712418}}].

\bibitem{Weinzierl:1998we}
S.~Weinzierl, {\it Reduction of multi-leg loop integrals},  {\em Phys. Lett.}
  {\bf B450} (1999) 234, [\href{http://xxx.lanl.gov/abs/hep-ph/9811365}{{\tt
  hep-ph/9811365}}].

\bibitem{delAguila:2004nf}
F.~del Aguila and R.~Pittau, {\it Recursive numerical calculus of one-loop
  tensor integrals},  {\em JHEP} {\bf 07} (2004) 017,
  [\href{http://xxx.lanl.gov/abs/hep-ph/0404120}{{\tt hep-ph/0404120}}].

\bibitem{vanHameren:2005ed}
A.~van Hameren, J.~Vollinga, and S.~Weinzierl, {\it Automated computation of
  one-loop integrals in massless theories},  {\em Eur. Phys. J.} {\bf C41}
  (2005) 361--375, [\href{http://xxx.lanl.gov/abs/hep-ph/0502165}{{\tt
  hep-ph/0502165}}].

\bibitem{Ossola:2006us}
G.~Ossola, C.~G. Papadopoulos, and R.~Pittau, {\it Reducing full one-loop
  amplitudes to scalar integrals at the integrand level},  {\em Nucl. Phys.}
  {\bf B763} (2007) 147--169,
  [\href{http://xxx.lanl.gov/abs/hep-ph/0609007}{{\tt hep-ph/0609007}}].

\bibitem{Bern:1994zx}
Z.~Bern, L.~Dixon, D.~C. Dunbar, and D.~A. Kosower, {\it One loop n point gauge
  theory amplitudes, unitarity and collinear limits},  {\em Nucl. Phys.} {\bf
  B425} (1994) 217--260, [\href{http://xxx.lanl.gov/abs/hep-ph/9403226}{{\tt
  hep-ph/9403226}}].

\bibitem{Bern:1995cg}
Z.~Bern, L.~Dixon, D.~C. Dunbar, and D.~A. Kosower, {\it Fusing gauge theory
  tree amplitudes into loop amplitudes},  {\em Nucl. Phys.} {\bf B435} (1995)
  59--101, [\href{http://xxx.lanl.gov/abs/hep-ph/9409265}{{\tt
  hep-ph/9409265}}].

\bibitem{Bern:1997ka}
Z.~Bern, L.~Dixon, D.~A. Kosower, and S.~Weinzierl, {\it One-loop amplitudes
  for e+ e- --> anti-q q anti-q q},  {\em Nucl. Phys.} {\bf B489} (1997) 3--23,
  [\href{http://xxx.lanl.gov/abs/hep-ph/9610370}{{\tt hep-ph/9610370}}].

\bibitem{Bern:1997sc}
Z.~Bern, L.~Dixon, and D.~A. Kosower, {\it One-loop amplitudes for e+ e- to
  four partons},  {\em Nucl. Phys.} {\bf B513} (1998) 3,
  [\href{http://xxx.lanl.gov/abs/hep-ph/9708239}{{\tt hep-ph/9708239}}].

\bibitem{Britto:2004nc}
R.~Britto, F.~Cachazo, and B.~Feng, {\it Generalized unitarity and one-loop
  amplitudes in n = 4 super-yang-mills},  {\em Nucl. Phys.} {\bf B725} (2005)
  275--305, [\href{http://xxx.lanl.gov/abs/hep-th/0412103}{{\tt
  hep-th/0412103}}].

\bibitem{Bidder:2004tx}
S.~J. Bidder, N.~E.~J. Bjerrum-Bohr, L.~J. Dixon, and D.~C. Dunbar, {\it N = 1
  supersymmetric one-loop amplitudes and the holomorphic anomaly of unitarity
  cuts},  {\em Phys. Lett.} {\bf B606} (2005) 189--201,
  [\href{http://xxx.lanl.gov/abs/hep-th/0410296}{{\tt hep-th/0410296}}].

\bibitem{Bidder:2004vx}
S.~J. Bidder, N.~E.~J. Bjerrum-Bohr, D.~C. Dunbar, and W.~B. Perkins, {\it
  Twistor space structure of the box coefficients of n = 1 one-loop
  amplitudes},  {\em Phys. Lett.} {\bf B608} (2005) 151--163,
  [\href{http://xxx.lanl.gov/abs/hep-th/0412023}{{\tt hep-th/0412023}}].

\bibitem{Bidder:2005ri}
S.~J. Bidder, N.~E.~J. Bjerrum-Bohr, D.~C. Dunbar, and W.~B. Perkins, {\it
  One-loop gluon scattering amplitudes in theories with n < 4 supersymmetries},
   {\em Phys. Lett.} {\bf B612} (2005) 75--88,
  [\href{http://xxx.lanl.gov/abs/hep-th/0502028}{{\tt hep-th/0502028}}].

\bibitem{Bedford:2004nh}
J.~Bedford, A.~Brandhuber, B.~J. Spence, and G.~Travaglini, {\it
  Non-supersymmetric loop amplitudes and mhv vertices},  {\em Nucl. Phys.} {\bf
  B712} (2005) 59--85, [\href{http://xxx.lanl.gov/abs/hep-th/0412108}{{\tt
  hep-th/0412108}}].

\bibitem{Britto:2005ha}
R.~Britto, E.~Buchbinder, F.~Cachazo, and B.~Feng, {\it One-loop amplitudes of
  gluons in sqcd},  {\em Phys. Rev.} {\bf D72} (2005) 065012,
  [\href{http://xxx.lanl.gov/abs/hep-ph/0503132}{{\tt hep-ph/0503132}}].

\bibitem{Bern:2005hs}
Z.~Bern, L.~J. Dixon, and D.~A. Kosower, {\it On-shell recurrence relations for
  one-loop qcd amplitudes},  {\em Phys. Rev.} {\bf D71} (2005) 105013,
  [\href{http://xxx.lanl.gov/abs/hep-th/0501240}{{\tt hep-th/0501240}}].

\bibitem{Bern:2005ji}
Z.~Bern, L.~J. Dixon, and D.~A. Kosower, {\it The last of the finite loop
  amplitudes in qcd},  {\em Phys. Rev.} {\bf D72} (2005) 125003,
  [\href{http://xxx.lanl.gov/abs/hep-ph/0505055}{{\tt hep-ph/0505055}}].

\bibitem{Bern:2005cq}
Z.~Bern, L.~J. Dixon, and D.~A. Kosower, {\it Bootstrapping multi-parton loop
  amplitudes in qcd},  {\em Phys. Rev.} {\bf D73} (2006) 065013,
  [\href{http://xxx.lanl.gov/abs/hep-ph/0507005}{{\tt hep-ph/0507005}}].

\bibitem{Forde:2005hh}
D.~Forde and D.~A. Kosower, {\it All-multiplicity one-loop corrections to mhv
  amplitudes in qcd},  {\em Phys. Rev.} {\bf D73} (2006) 061701,
  [\href{http://xxx.lanl.gov/abs/hep-ph/0509358}{{\tt hep-ph/0509358}}].

\bibitem{Berger:2006ci}
C.~F. Berger, Z.~Bern, L.~J. Dixon, D.~Forde, and D.~A. Kosower, {\it
  Bootstrapping one-loop qcd amplitudes with general helicities},  {\em Phys.
  Rev.} {\bf D74} (2006) 036009,
  [\href{http://xxx.lanl.gov/abs/hep-ph/0604195}{{\tt hep-ph/0604195}}].

\bibitem{Berger:2006vq}
C.~F. Berger, Z.~Bern, L.~J. Dixon, D.~Forde, and D.~A. Kosower, {\it All
  one-loop maximally helicity violating gluonic amplitudes in qcd},  {\em Phys.
  Rev.} {\bf D75} (2007) 016006,
  [\href{http://xxx.lanl.gov/abs/hep-ph/0607014}{{\tt hep-ph/0607014}}].

\bibitem{Britto:2006sj}
R.~Britto, B.~Feng, and P.~Mastrolia, {\it The cut-constructible part of qcd
  amplitudes},  {\em Phys. Rev.} {\bf D73} (2006) 105004,
  [\href{http://xxx.lanl.gov/abs/hep-ph/0602178}{{\tt hep-ph/0602178}}].

\bibitem{Anastasiou:2006jv}
C.~Anastasiou, R.~Britto, B.~Feng, Z.~Kunszt, and P.~Mastrolia, {\it
  D-dimensional unitarity cut method},  {\em Phys. Lett.} {\bf B645} (2007)
  213--216, [\href{http://xxx.lanl.gov/abs/hep-ph/0609191}{{\tt
  hep-ph/0609191}}].

\bibitem{Anastasiou:2006gt}
C.~Anastasiou, R.~Britto, B.~Feng, Z.~Kunszt, and P.~Mastrolia, {\it Unitarity
  cuts and reduction to master integrals in d dimensions for one-loop
  amplitudes},  {\em JHEP} {\bf 03} (2007) 111,
  [\href{http://xxx.lanl.gov/abs/hep-ph/0612277}{{\tt hep-ph/0612277}}].

\bibitem{Mastrolia:2006ki}
P.~Mastrolia, {\it On triple-cut of scattering amplitudes},  {\em Phys. Lett.}
  {\bf B644} (2007) 272--283,
  [\href{http://xxx.lanl.gov/abs/hep-th/0611091}{{\tt hep-th/0611091}}].

\bibitem{Britto:2006fc}
R.~Britto and B.~Feng, {\it Unitarity cuts with massive propagators and
  algebraic expressions for coefficients},  {\em Phys. Rev.} {\bf D75} (2007)
  105006, [\href{http://xxx.lanl.gov/abs/hep-ph/0612089}{{\tt
  hep-ph/0612089}}].

\bibitem{Forde:2007mi}
D.~Forde, {\it Direct extraction of one-loop integral coefficients},  {\em
  Phys. Rev.} {\bf D75} (2007) 125019,
  [\href{http://xxx.lanl.gov/abs/arXiv:0704.1835 [hep-ph]}{{\tt arXiv:0704.1835
  [hep-ph]}}].

\bibitem{Bern:2005hh}
Z.~Bern, N.~E.~J. Bjerrum-Bohr, D.~C. Dunbar, and H.~Ita, {\it Recursive
  calculation of one-loop qcd integral coefficients},  {\em JHEP} {\bf 11}
  (2005) 027, [\href{http://xxx.lanl.gov/abs/hep-ph/0507019}{{\tt
  hep-ph/0507019}}].

\bibitem{Xiao:2006vr}
Z.~Xiao, G.~Yang, and C.-J. Zhu, {\it The rational part of qcd amplitude. i:
  The general formalism},  {\em Nucl. Phys.} {\bf B758} (2006) 1--34,
  [\href{http://xxx.lanl.gov/abs/hep-ph/0607015}{{\tt hep-ph/0607015}}].

\bibitem{Su:2006vs}
X.~Su, Z.~Xiao, G.~Yang, and C.-J. Zhu, {\it The rational part of qcd
  amplitude. ii: The five-gluon},  {\em Nucl. Phys.} {\bf B758} (2006) 35--52,
  [\href{http://xxx.lanl.gov/abs/hep-ph/0607016}{{\tt hep-ph/0607016}}].

\bibitem{Xiao:2006vt}
Z.~Xiao, G.~Yang, and C.-J. Zhu, {\it The rational part of qcd amplitude. iii:
  The six-gluon},  {\em Nucl. Phys.} {\bf B758} (2006) 53--89,
  [\href{http://xxx.lanl.gov/abs/hep-ph/0607017}{{\tt hep-ph/0607017}}].

\bibitem{Ellis:2006ss}
R.~K. Ellis, W.~T. Giele, and G.~Zanderighi, {\it The one-loop amplitude for
  six-gluon scattering},  {\em JHEP} {\bf 05} (2006) 027,
  [\href{http://xxx.lanl.gov/abs/hep-ph/0602185}{{\tt hep-ph/0602185}}].

\bibitem{Binoth:2006hk}
T.~Binoth, J.~P. Guillet, and G.~Heinrich, {\it Algebraic evaluation of
  rational polynomials in one-loop amplitudes},  {\em JHEP} {\bf 02} (2007)
  013, [\href{http://xxx.lanl.gov/abs/hep-ph/0609054}{{\tt hep-ph/0609054}}].

\bibitem{Nagy:2006xy}
Z.~Nagy and D.~E. Soper, {\it Numerical integration of one-loop feynman
  diagrams for n- photon amplitudes},  {\em Phys. Rev.} {\bf D74} (2006)
  093006, [\href{http://xxx.lanl.gov/abs/hep-ph/0610028}{{\tt
  hep-ph/0610028}}].

\bibitem{Binoth:2007ca}
T.~Binoth, G.~Heinrich, T.~Gehrmann, and P.~Mastrolia, {\it Six-photon
  amplitudes},  {\em Phys. Lett.} {\bf B649} (2007) 422--426,
  [\href{http://xxx.lanl.gov/abs/hep-ph/0703311}{{\tt hep-ph/0703311}}].

\bibitem{Ossola:2007bb}
G.~Ossola, C.~G. Papadopoulos, and R.~Pittau, {\it Numerical evaluation of
  six-photon amplitudes},  \href{http://xxx.lanl.gov/abs/arXiv:0704.1271
  [hep-ph]}{{\tt arXiv:0704.1271 [hep-ph]}}.

\bibitem{Chetyrkin:1981qh}
K.~G. Chetyrkin and F.~V. Tkachov, {\it Integration by parts: The algorithm to
  calculate beta functions in 4 loops},  {\em Nucl. Phys.} {\bf B192} (1981)
  159--204.

\bibitem{Tarasov:1996br}
O.~V. Tarasov, {\it Connection between feynman integrals having different
  values of the space-time dimension},  {\em Phys. Rev.} {\bf D54} (1996)
  6479--6490, [\href{http://xxx.lanl.gov/abs/hep-th/9606018}{{\tt
  hep-th/9606018}}].

\bibitem{Tarasov:1997kx}
O.~V. Tarasov, {\it Generalized recurrence relations for two-loop propagator
  integrals with arbitrary masses},  {\em Nucl. Phys.} {\bf B502} (1997)
  455--482, [\href{http://xxx.lanl.gov/abs/hep-ph/9703319}{{\tt
  hep-ph/9703319}}].

\bibitem{Laporta:2001dd}
S.~Laporta, {\it High-precision calculation of multi-loop feynman integrals by
  difference equations},  {\em Int. J. Mod. Phys.} {\bf A15} (2000) 5087--5159,
  [\href{http://xxx.lanl.gov/abs/hep-ph/0102033}{{\tt hep-ph/0102033}}].

\bibitem{Bern:2000ie}
Z.~Bern, L.~Dixon, and A.~Ghinculov, {\it Two-loop correction to bhabha
  scattering},  {\em Phys. Rev.} {\bf D63} (2001) 053007,
  [\href{http://xxx.lanl.gov/abs/hep-ph/0010075}{{\tt hep-ph/0010075}}].

\bibitem{Bern:2000dn}
Z.~Bern, L.~Dixon, and D.~A. Kosower, {\it A two-loop four-gluon helicity
  amplitude in qcd},  {\em JHEP} {\bf 01} (2000) 027,
  [\href{http://xxx.lanl.gov/abs/hep-ph/0001001}{{\tt hep-ph/0001001}}].

\bibitem{Anastasiou:2000kg}
C.~Anastasiou, E.~W.~N. Glover, C.~Oleari, and M.~E. Tejeda-Yeomans, {\it
  Two-loop qcd corrections to q anti-q --> q' anti-q'},  {\em Nucl. Phys.} {\bf
  B601} (2001) 318--340, [\href{http://xxx.lanl.gov/abs/hep-ph/0010212}{{\tt
  hep-ph/0010212}}].

\bibitem{Anastasiou:2000ue}
C.~Anastasiou, E.~W.~N. Glover, C.~Oleari, and M.~E. Tejeda-Yeomans, {\it
  Two-loop qcd corrections to q anti-q --> q anti-q},  {\em Nucl. Phys.} {\bf
  B601} (2001) 341--360, [\href{http://xxx.lanl.gov/abs/hep-ph/0011094}{{\tt
  hep-ph/0011094}}].

\bibitem{Anastasiou:2000mv}
C.~Anastasiou, E.~W.~N. Glover, C.~Oleari, and M.~E. Tejeda-Yeomans, {\it
  One-loop qcd corrections to quark scattering at nnlo},  {\em Phys. Lett.}
  {\bf B506} (2001) 59--67, [\href{http://xxx.lanl.gov/abs/hep-ph/0012007}{{\tt
  hep-ph/0012007}}].

\bibitem{Anastasiou:2001sv}
C.~Anastasiou, E.~W.~N. Glover, C.~Oleari, and M.~E. Tejeda-Yeomans, {\it
  Two-loop qcd corrections to massless quark gluon scattering},  {\em Nucl.
  Phys.} {\bf B605} (2001) 486--516,
  [\href{http://xxx.lanl.gov/abs/hep-ph/0101304}{{\tt hep-ph/0101304}}].

\bibitem{Glover:2001af}
E.~W.~N. Glover, C.~Oleari, and M.~E. Tejeda-Yeomans, {\it Two-loop qcd
  corrections to gluon gluon scattering},  {\em Nucl. Phys.} {\bf B605} (2001)
  467--485, [\href{http://xxx.lanl.gov/abs/hep-ph/0102201}{{\tt
  hep-ph/0102201}}].

\bibitem{Bern:2002tk}
Z.~Bern, A.~De~Freitas, and L.~Dixon, {\it Two-loop helicity amplitudes for
  gluon gluon scattering in qcd and supersymmetric yang-mills theory},  {\em
  JHEP} {\bf 03} (2002) 018,
  [\href{http://xxx.lanl.gov/abs/hep-ph/0201161}{{\tt hep-ph/0201161}}].

\bibitem{Garland:2001tf}
L.~W. Garland, T.~Gehrmann, E.~W.~N. Glover, A.~Koukoutsakis, and E.~Remiddi,
  {\it The two-loop qcd matrix element for e+ e- --> 3jets},  {\em Nucl. Phys.}
  {\bf B627} (2002) 107--188,
  [\href{http://xxx.lanl.gov/abs/hep-ph/0112081}{{\tt hep-ph/0112081}}].

\bibitem{Garland:2002ak}
L.~W. Garland, T.~Gehrmann, E.~W.~N. Glover, A.~Koukoutsakis, and E.~Remiddi,
  {\it Two-loop qcd helicity amplitudes for e+ e- --> 3jets},  {\em Nucl.
  Phys.} {\bf B642} (2002) 227--262,
  [\href{http://xxx.lanl.gov/abs/hep-ph/0206067}{{\tt hep-ph/0206067}}].

\bibitem{Moch:2002hm}
S.~Moch, P.~Uwer, and S.~Weinzierl, {\it Two-loop amplitudes with nested sums:
  Fermionic contributions to e+ e- --> q anti-q g},  {\em Phys. Rev.} {\bf D66}
  (2002) 114001, [\href{http://xxx.lanl.gov/abs/hep-ph/0207043}{{\tt
  hep-ph/0207043}}].

\bibitem{Harlander:2000mg}
R.~V. Harlander, {\it Virtual corrections to g g --> h to two loops in the
  heavy top limit},  {\em Phys. Lett.} {\bf B492} (2000) 74--80,
  [\href{http://xxx.lanl.gov/abs/hep-ph/0007289}{{\tt hep-ph/0007289}}].

\bibitem{Ravindran:2004mb}
V.~Ravindran, J.~Smith, and W.~L. van Neerven, {\it Two-loop corrections to
  higgs boson production},  {\em Nucl. Phys.} {\bf B704} (2005) 332--348,
  [\href{http://xxx.lanl.gov/abs/hep-ph/0408315}{{\tt hep-ph/0408315}}].

\bibitem{Moch:2004pa}
S.~Moch, J.~A.~M. Vermaseren, and A.~Vogt, {\it The three-loop splitting
  functions in qcd: The non-singlet case},  {\em Nucl. Phys.} {\bf B688} (2004)
  101--134, [\href{http://xxx.lanl.gov/abs/hep-ph/0403192}{{\tt
  hep-ph/0403192}}].

\bibitem{Vogt:2004mw}
A.~Vogt, S.~Moch, and J.~A.~M. Vermaseren, {\it The three-loop splitting
  functions in qcd: The singlet case},  {\em Nucl. Phys.} {\bf B691} (2004)
  129--181, [\href{http://xxx.lanl.gov/abs/hep-ph/0404111}{{\tt
  hep-ph/0404111}}].

\bibitem{Boos:1990rg}
E.~E. Boos and A.~I. Davydychev, {\it A method of evaluating massive feynman
  integrals},  {\em Theor. Math. Phys.} {\bf 89} (1991) 1052--1063.

\bibitem{Davydychev:1990jt}
A.~I. Davydychev, {\it Some exact results for n point massive feynman
  integrals},  {\em J. Math. Phys.} {\bf 32} (1991) 1052--1060.

\bibitem{Davydychev:1990cq}
A.~I. Davydychev, {\it General results for massive n point feynman diagrams
  with different masses},  {\em J. Math. Phys.} {\bf 33} (1992) 358--369.

\bibitem{Smirnov:1999gc}
V.~A. Smirnov, {\it Analytical result for dimensionally regularized massless
  on-shell double box},  {\em Phys. Lett.} {\bf B460} (1999) 397--404,
  [\href{http://xxx.lanl.gov/abs/hep-ph/9905323}{{\tt hep-ph/9905323}}].

\bibitem{Smirnov:1999wz}
V.~A. Smirnov and O.~L. Veretin, {\it Analytical results for dimensionally
  regularized massless on-shell double boxes with arbitrary indices and
  numerators},  {\em Nucl. Phys.} {\bf B566} (2000) 469--485,
  [\href{http://xxx.lanl.gov/abs/hep-ph/9907385}{{\tt hep-ph/9907385}}].

\bibitem{Tausk:1999vh}
J.~B. Tausk, {\it Non-planar massless two-loop feynman diagrams with four on-
  shell legs},  {\em Phys. Lett.} {\bf B469} (1999) 225--234,
  [\href{http://xxx.lanl.gov/abs/hep-ph/9909506}{{\tt hep-ph/9909506}}].

\bibitem{Smirnov:2000vy}
V.~A. Smirnov, {\it Analytical result for dimensionally regularized massless
  master double box with one leg off shell},  {\em Phys. Lett.} {\bf B491}
  (2000) 130--136, [\href{http://xxx.lanl.gov/abs/hep-ph/0007032}{{\tt
  hep-ph/0007032}}].

\bibitem{Smirnov:2000ie}
V.~A. Smirnov, {\it Analytical result for dimensionally regularized massless
  master non-planar double box with one leg off shell},  {\em Phys. Lett.} {\bf
  B500} (2001) 330--337, [\href{http://xxx.lanl.gov/abs/hep-ph/0011056}{{\tt
  hep-ph/0011056}}].

\bibitem{Smirnov:2003vi}
V.~A. Smirnov, {\it Analytical result for dimensionally regularized massless
  on-shell planar triple box},  {\em Phys. Lett.} {\bf B567} (2003) 193--199,
  [\href{http://xxx.lanl.gov/abs/hep-ph/0305142}{{\tt hep-ph/0305142}}].

\bibitem{Bierenbaum:2003ud}
I.~Bierenbaum and S.~Weinzierl, {\it The massless two-loop two-point function},
   {\em Eur. Phys. J.} {\bf C32} (2003) 67--78,
  [\href{http://xxx.lanl.gov/abs/hep-ph/0308311}{{\tt hep-ph/0308311}}].

\bibitem{Heinrich:2004iq}
G.~Heinrich and V.~A. Smirnov, {\it Analytical evaluation of dimensionally
  regularized massive on-shell double boxes},  {\em Phys. Lett.} {\bf B598}
  (2004) 55--66, [\href{http://xxx.lanl.gov/abs/hep-ph/0406053}{{\tt
  hep-ph/0406053}}].

\bibitem{Friot:2005cu}
S.~Friot, D.~Greynat, and E.~De~Rafael, {\it Asymptotics of feynman diagrams
  and the mellin-barnes representation},  {\em Phys. Lett.} {\bf B628} (2005)
  73--84, [\href{http://xxx.lanl.gov/abs/hep-ph/0505038}{{\tt
  hep-ph/0505038}}].

\bibitem{Bern:2005iz}
Z.~Bern, L.~J. Dixon, and V.~A. Smirnov, {\it Iteration of planar amplitudes in
  maximally supersymmetric yang-mills theory at three loops and beyond},  {\em
  Phys. Rev.} {\bf D72} (2005) 085001,
  [\href{http://xxx.lanl.gov/abs/hep-th/0505205}{{\tt hep-th/0505205}}].

\bibitem{Anastasiou:2005cb}
C.~Anastasiou and A.~Daleo, {\it Numerical evaluation of loop integrals},  {\em
  JHEP} {\bf 10} (2006) 031,
  [\href{http://xxx.lanl.gov/abs/hep-ph/0511176}{{\tt hep-ph/0511176}}].

\bibitem{Czakon:2005rk}
M.~Czakon, {\it Automatized analytic continuation of mellin-barnes integrals},
  {\em Comput. Phys. Commun.} {\bf 175} (2006) 559--571,
  [\href{http://xxx.lanl.gov/abs/hep-ph/0511200}{{\tt hep-ph/0511200}}].

\bibitem{Gluza:2007rt}
J.~Gluza, K.~Kajda, and T.~Riemann, {\it Ambre - a mathematica package for the
  construction of mellin-barnes representations for feynman integrals},
  \href{http://xxx.lanl.gov/abs/arXiv:0704.2423 [hep-ph]}{{\tt arXiv:0704.2423
  [hep-ph]}}.

\bibitem{Hain}
R.~M. Hain, {\it Classical polylogarithms},
  \href{http://xxx.lanl.gov/abs/alg-geom/9202022}{{\tt alg-geom/9202022}}.

\bibitem{Goncharov}
A.~B. Goncharov, {\it Multiple polylogarithms, cyclotomy and modular
  complexes},  {\em Math. Res. Lett.} {\bf 5} (1998) 497,
  [\href{http://xxx.lanl.gov/abs/(available at
  http://www.math.uiuc.edu/K-theory/0297)}{{\tt (available at
  http://www.math.uiuc.edu/K-theory/0297)}}].

\bibitem{Borwein}
J.~M. Borwein, D.~M. Bradley, D.~J. Broadhurst, and P.~Lisonek, {\it Special
  values of multiple polylogarithms},  {\em Trans. Amer. Math. Soc.} {\bf
  353:3} (2001) 907, [\href{http://xxx.lanl.gov/abs/math.CA/9910045}{{\tt
  math.CA/9910045}}].

\bibitem{Minh:2000}
H.~M. Minh, M.~Petitot, and J.~van~der Hoeven, {\it Shuffle algebra and
  polylogarithms},  {\em Discrete Math.} {\bf 225:1-3} (2000) 217.

\bibitem{Remiddi:1999ew}
E.~Remiddi and J.~A.~M. Vermaseren, {\it Harmonic polylogarithms},  {\em Int.
  J. Mod. Phys.} {\bf A15} (2000) 725,
  [\href{http://xxx.lanl.gov/abs/hep-ph/9905237}{{\tt hep-ph/9905237}}].

\bibitem{Vermaseren:1998uu}
J.~A.~M. Vermaseren, {\it Harmonic sums, mellin transforms and integrals},
  {\em Int. J. Mod. Phys.} {\bf A14} (1999) 2037,
  [\href{http://xxx.lanl.gov/abs/hep-ph/9806280}{{\tt hep-ph/9806280}}].

\bibitem{Gehrmann:2000zt}
T.~Gehrmann and E.~Remiddi, {\it Two-loop master integrals for gamma* -->
  3jets: The planar topologies},  {\em Nucl. Phys.} {\bf B601} (2001) 248--286,
  [\href{http://xxx.lanl.gov/abs/hep-ph/0008287}{{\tt hep-ph/0008287}}].

\bibitem{Gehrmann:2001pz}
T.~Gehrmann and E.~Remiddi, {\it Numerical evaluation of harmonic
  polylogarithms},  {\em Comput. Phys. Commun.} {\bf 141} (2001) 296--312,
  [\href{http://xxx.lanl.gov/abs/hep-ph/0107173}{{\tt hep-ph/0107173}}].

\bibitem{Gehrmann:2001jv}
T.~Gehrmann and E.~Remiddi, {\it Numerical evaluation of two-dimensional
  harmonic polylogarithms},  {\em Comput. Phys. Commun.} {\bf 144} (2002)
  200--223, [\href{http://xxx.lanl.gov/abs/hep-ph/0111255}{{\tt
  hep-ph/0111255}}].

\bibitem{Gehrmann:2002zr}
T.~Gehrmann and E.~Remiddi, {\it Analytic continuation of massless two-loop
  four-point functions},  {\em Nucl. Phys.} {\bf B640} (2002) 379--411,
  [\href{http://xxx.lanl.gov/abs/hep-ph/0207020}{{\tt hep-ph/0207020}}].

\bibitem{Moch:2001zr}
S.~Moch, P.~Uwer, and S.~Weinzierl, {\it Nested sums, expansion of
  transcendental functions and multi-scale multi-loop integrals},  {\em J.
  Math. Phys.} {\bf 43} (2002) 3363--3386,
  [\href{http://xxx.lanl.gov/abs/hep-ph/0110083}{{\tt hep-ph/0110083}}].

\bibitem{Blumlein:1998if}
J.~Bl{\"u}mlein and S.~Kurth, {\it Harmonic sums and mellin transforms up to
  two-loop order},  {\em Phys. Rev.} {\bf D60} (1999) 014018,
  [\href{http://xxx.lanl.gov/abs/hep-ph/9810241}{{\tt hep-ph/9810241}}].

\bibitem{Blumlein:2003gb}
J.~Bl{\"u}mlein, {\it Algebraic relations between harmonic sums and associated
  quantities},  {\em Comput. Phys. Commun.} {\bf 159} (2004) 19--54,
  [\href{http://xxx.lanl.gov/abs/hep-ph/0311046}{{\tt hep-ph/0311046}}].

\bibitem{Weinzierl:2004bn}
S.~Weinzierl, {\it Expansion around half-integer values, binomial sums and
  inverse binomial sums},  {\em J. Math. Phys.} {\bf 45} (2004) 2656--2673,
  [\href{http://xxx.lanl.gov/abs/hep-ph/0402131}{{\tt hep-ph/0402131}}].

\bibitem{Vollinga:2004sn}
J.~Vollinga and S.~Weinzierl, {\it Numerical evaluation of multiple
  polylogarithms},  {\em Comput. Phys. Commun.} {\bf 167} (2005) 177,
  [\href{http://xxx.lanl.gov/abs/hep-ph/0410259}{{\tt hep-ph/0410259}}].

\bibitem{Korner:2005qz}
J.~G. K{\"o}rner, Z.~Merebashvili, and M.~Rogal, {\it Laurent series expansion
  of a class of massive scalar one- loop integrals up to o(epsilon**2) in terms
  of multiple polylogarithms},  {\em J. Math. Phys.} {\bf 47} (2006) 072302,
  [\href{http://xxx.lanl.gov/abs/hep-ph/0512159}{{\tt hep-ph/0512159}}].

\bibitem{Kalmykov:2006hu}
M.~Y. Kalmykov, B.~F.~L. Ward, and S.~Yost, {\it All order epsilon-expansion of
  gauss hypergeometric functions with integer and half/integer values of
  parameters},  {\em JHEP} {\bf 02} (2007) 040,
  [\href{http://xxx.lanl.gov/abs/hep-th/0612240}{{\tt hep-th/0612240}}].

\bibitem{Maitre:2007kp}
D.~Maitre, {\it Extension of hpl to complex arguments},
  \href{http://xxx.lanl.gov/abs/hep-ph/0703052}{{\tt hep-ph/0703052}}.

\bibitem{Weinzierl:2002hv}
S.~Weinzierl, {\it Symbolic expansion of transcendental functions},  {\em
  Comput. Phys. Commun.} {\bf 145} (2002) 357--370,
  [\href{http://xxx.lanl.gov/abs/math-ph/0201011}{{\tt math-ph/0201011}}].

\bibitem{Moch:2005uc}
S.~Moch and P.~Uwer, {\it Xsummer: Transcendental functions and symbolic
  summation in form},  {\em Comput. Phys. Commun.} {\bf 174} (2006) 759--770,
  [\href{http://xxx.lanl.gov/abs/math-ph/0508008}{{\tt math-ph/0508008}}].

\bibitem{Maitre:2005uu}
D.~Maitre, {\it Hpl, a mathematica implementation of the harmonic
  polylogarithms},  {\em Comput. Phys. Commun.} {\bf 174} (2006) 222--240,
  [\href{http://xxx.lanl.gov/abs/hep-ph/0507152}{{\tt hep-ph/0507152}}].

\bibitem{Huber:2005yg}
T.~Huber and D.~Maitre, {\it Hypexp, a mathematica package for expanding
  hypergeometric functions around integer-valued parameters},  {\em Comput.
  Phys. Commun.} {\bf 175} (2006) 122--144,
  [\href{http://xxx.lanl.gov/abs/hep-ph/0507094}{{\tt hep-ph/0507094}}].

\bibitem{'tHooft:1979xw}
G.~'t~Hooft and M.~J.~G. Veltman, {\it Scalar one loop integrals},  {\em Nucl.
  Phys.} {\bf B153} (1979) 365--401.

\bibitem{Hepp:1966eg}
K.~Hepp, {\it Proof of the bogolyubov-parasiuk theorem on renormalization},
  {\em Commun. Math. Phys.} {\bf 2} (1966) 301--326.

\bibitem{Roth:1996pd}
M.~Roth and A.~Denner, {\it High-energy approximation of one-loop feynman
  integrals},  {\em Nucl. Phys.} {\bf B479} (1996) 495--514,
  [\href{http://xxx.lanl.gov/abs/hep-ph/9605420}{{\tt hep-ph/9605420}}].

\bibitem{Binoth:2000ps}
T.~Binoth and G.~Heinrich, {\it An automatized algorithm to compute infrared
  divergent multi-loop integrals},  {\em Nucl. Phys.} {\bf B585} (2000)
  741--759, [\href{http://xxx.lanl.gov/abs/hep-ph/0004013}{{\tt
  hep-ph/0004013}}].

\bibitem{Kinoshita:1962ur}
T.~Kinoshita, {\it Mass singularities of feynman amplitudes},  {\em J. Math.
  Phys.} {\bf 3} (1962) 650--677.

\bibitem{Lee:1964is}
T.~D. Lee and M.~Nauenberg, {\it Degenerate systems and mass singularities},
  {\em Phys. Rev.} {\bf 133} (1964) B1549--B1562.

\bibitem{Giele:1992vf}
W.~T. Giele and E.~W.~N. Glover, {\it Higher order corrections to jet
  cross-sections in e+ e- annihilation},  {\em Phys. Rev.} {\bf D46} (1992)
  1980--2010.

\bibitem{Giele:1993dj}
W.~T. Giele, E.~W.~N. Glover, and D.~A. Kosower, {\it Higher order corrections
  to jet cross-sections in hadron colliders},  {\em Nucl. Phys.} {\bf B403}
  (1993) 633--670, [\href{http://xxx.lanl.gov/abs/hep-ph/9302225}{{\tt
  hep-ph/9302225}}].

\bibitem{Keller:1998tf}
S.~Keller and E.~Laenen, {\it Next-to-leading order cross sections for tagged
  reactions},  {\em Phys. Rev.} {\bf D59} (1999) 114004,
  [\href{http://xxx.lanl.gov/abs/hep-ph/9812415}{{\tt hep-ph/9812415}}].

\bibitem{Frixione:1996ms}
S.~Frixione, Z.~Kunszt, and A.~Signer, {\it Three jet cross-sections to
  next-to-leading order},  {\em Nucl. Phys.} {\bf B467} (1996) 399--442,
  [\href{http://xxx.lanl.gov/abs/hep-ph/9512328}{{\tt hep-ph/9512328}}].

\bibitem{Catani:1997vz}
S.~Catani and M.~H. Seymour, {\it A general algorithm for calculating jet
  cross-sections in nlo qcd},  {\em Nucl. Phys.} {\bf B485} (1997) 291--419,
  [\href{http://xxx.lanl.gov/abs/hep-ph/9605323}{{\tt hep-ph/9605323}}].

\bibitem{Dittmaier:1999mb}
S.~Dittmaier, {\it A general approach to photon radiation off fermions},  {\em
  Nucl. Phys.} {\bf B565} (2000) 69--122,
  [\href{http://xxx.lanl.gov/abs/hep-ph/9904440}{{\tt hep-ph/9904440}}].

\bibitem{Phaf:2001gc}
L.~Phaf and S.~Weinzierl, {\it Dipole formalism with heavy fermions},  {\em
  JHEP} {\bf 04} (2001) 006,
  [\href{http://xxx.lanl.gov/abs/hep-ph/0102207}{{\tt hep-ph/0102207}}].

\bibitem{Catani:2002hc}
S.~Catani, S.~Dittmaier, M.~H. Seymour, and Z.~Trocsanyi, {\it The dipole
  formalism for next-to-leading order qcd calculations with massive partons},
  {\em Nucl. Phys.} {\bf B627} (2002) 189--265,
  [\href{http://xxx.lanl.gov/abs/hep-ph/0201036}{{\tt hep-ph/0201036}}].

\bibitem{Kosower:1998zr}
D.~A. Kosower, {\it Antenna factorization of gauge-theory amplitudes},  {\em
  Phys. Rev.} {\bf D57} (1998) 5410--5416,
  [\href{http://xxx.lanl.gov/abs/hep-ph/9710213}{{\tt hep-ph/9710213}}].

\bibitem{Daleo:2006xa}
A.~Daleo, T.~Gehrmann, and D.~Maitre, {\it Antenna subtraction with hadronic
  initial states},  {\em JHEP} {\bf 04} (2007) 016,
  [\href{http://xxx.lanl.gov/abs/hep-ph/0612257}{{\tt hep-ph/0612257}}].

\bibitem{Kosower:2002su}
D.~A. Kosower, {\it Multiple singular emission in gauge theories},  {\em Phys.
  Rev.} {\bf D67} (2003) 116003,
  [\href{http://xxx.lanl.gov/abs/hep-ph/0212097}{{\tt hep-ph/0212097}}].

\bibitem{Kosower:2003cz}
D.~A. Kosower, {\it All-orders singular emission in gauge theories},  {\em
  Phys. Rev. Lett.} {\bf 91} (2003) 061602,
  [\href{http://xxx.lanl.gov/abs/hep-ph/0301069}{{\tt hep-ph/0301069}}].

\bibitem{Kosower:2003bh}
D.~A. Kosower, {\it Antenna factorization in strongly-ordered limits},  {\em
  Phys. Rev.} {\bf D71} (2005) 045016,
  [\href{http://xxx.lanl.gov/abs/hep-ph/0311272}{{\tt hep-ph/0311272}}].

\bibitem{Weinzierl:2003fx}
S.~Weinzierl, {\it Subtraction terms at nnlo},  {\em JHEP} {\bf 03} (2003) 062,
  [\href{http://xxx.lanl.gov/abs/hep-ph/0302180}{{\tt hep-ph/0302180}}].

\bibitem{Weinzierl:2003ra}
S.~Weinzierl, {\it Subtraction terms for one-loop amplitudes with one
  unresolved parton},  {\em JHEP} {\bf 07} (2003) 052,
  [\href{http://xxx.lanl.gov/abs/hep-ph/0306248}{{\tt hep-ph/0306248}}].

\bibitem{Anastasiou:2003gr}
C.~Anastasiou, K.~Melnikov, and F.~Petriello, {\it A new method for real
  radiation at nnlo},  {\em Phys. Rev.} {\bf D69} (2004) 076010,
  [\href{http://xxx.lanl.gov/abs/hep-ph/0311311}{{\tt hep-ph/0311311}}].

\bibitem{Gehrmann-DeRidder:2003bm}
A.~Gehrmann-De~Ridder, T.~Gehrmann, and G.~Heinrich, {\it Four-particle phase
  space integrals in massless qcd},  {\em Nucl. Phys.} {\bf B682} (2004)
  265--288, [\href{http://xxx.lanl.gov/abs/hep-ph/0311276}{{\tt
  hep-ph/0311276}}].

\bibitem{Gehrmann-DeRidder:2004tv}
A.~Gehrmann-De~Ridder, T.~Gehrmann, and E.~W.~N. Glover, {\it Infrared
  structure of e+ e- --> 2jets at nnlo},  {\em Nucl. Phys.} {\bf B691} (2004)
  195--222, [\href{http://xxx.lanl.gov/abs/hep-ph/0403057}{{\tt
  hep-ph/0403057}}].

\bibitem{Gehrmann-DeRidder:2005hi}
A.~Gehrmann-De~Ridder, T.~Gehrmann, and E.~W.~N. Glover, {\it Quark-gluon
  antenna functions from neutralino decay},  {\em Phys. Lett.} {\bf B612}
  (2005) 36--48, [\href{http://xxx.lanl.gov/abs/hep-ph/0501291}{{\tt
  hep-ph/0501291}}].

\bibitem{Gehrmann-DeRidder:2005aw}
A.~Gehrmann-De~Ridder, T.~Gehrmann, and E.~W.~N. Glover, {\it Gluon gluon
  antenna functions from higgs boson decay},  {\em Phys. Lett.} {\bf B612}
  (2005) 49--60, [\href{http://xxx.lanl.gov/abs/hep-ph/0502110}{{\tt
  hep-ph/0502110}}].

\bibitem{Gehrmann-DeRidder:2005cm}
A.~Gehrmann-De~Ridder, T.~Gehrmann, and E.~W.~N. Glover, {\it Antenna
  subtraction at nnlo},  {\em JHEP} {\bf 09} (2005) 056,
  [\href{http://xxx.lanl.gov/abs/hep-ph/0505111}{{\tt hep-ph/0505111}}].

\bibitem{Binoth:2004jv}
T.~Binoth and G.~Heinrich, {\it Numerical evaluation of phase space integrals
  by sector decomposition},  {\em Nucl. Phys.} {\bf B693} (2004) 134--148,
  [\href{http://xxx.lanl.gov/abs/hep-ph/0402265}{{\tt hep-ph/0402265}}].

\bibitem{Heinrich:2006sw}
G.~Heinrich, {\it Towards e+ e- --> 3jets at nnlo by sector decomposition},
  {\em Eur. Phys. J.} {\bf C48} (2006) 25--33,
  [\href{http://xxx.lanl.gov/abs/hep-ph/0601062}{{\tt hep-ph/0601062}}].

\bibitem{Kilgore:2004ty}
W.~B. Kilgore, {\it Subtraction terms for hadronic production processes at
  next-to-next-to-leading order},  {\em Phys. Rev.} {\bf D70} (2004) 031501,
  [\href{http://xxx.lanl.gov/abs/hep-ph/0403128}{{\tt hep-ph/0403128}}].

\bibitem{Frixione:2004is}
S.~Frixione and M.~Grazzini, {\it Subtraction at nnlo},  {\em JHEP} {\bf 06}
  (2005) 010, [\href{http://xxx.lanl.gov/abs/hep-ph/0411399}{{\tt
  hep-ph/0411399}}].

\bibitem{Catani:2007vq}
S.~Catani and M.~Grazzini, {\it An nnlo subtraction formalism in hadron
  collisions and its application to higgs boson production at the lhc},  {\em
  Phys. Rev. Lett.} {\bf 98} (2007) 222002,
  [\href{http://xxx.lanl.gov/abs/hep-ph/0703012}{{\tt hep-ph/0703012}}].

\bibitem{Somogyi:2005xz}
G.~Somogyi, Z.~Trocsanyi, and V.~Del~Duca, {\it Matching of singly- and
  doubly-unresolved limits of tree- level qcd squared matrix elements},  {\em
  JHEP} {\bf 06} (2005) 024,
  [\href{http://xxx.lanl.gov/abs/hep-ph/0502226}{{\tt hep-ph/0502226}}].

\bibitem{Somogyi:2006da}
G.~Somogyi, Z.~Trocsanyi, and V.~Del~Duca, {\it A subtraction scheme for
  computing qcd jet cross sections at nnlo: Regularization of doubly-real
  emissions},  {\em JHEP} {\bf 01} (2007) 070,
  [\href{http://xxx.lanl.gov/abs/hep-ph/0609042}{{\tt hep-ph/0609042}}].

\bibitem{Somogyi:2006db}
G.~Somogyi and Z.~Trocsanyi, {\it A subtraction scheme for computing qcd jet
  cross sections at nnlo: Regularization of real-virtual emission},  {\em JHEP}
  {\bf 01} (2007) 052, [\href{http://xxx.lanl.gov/abs/hep-ph/0609043}{{\tt
  hep-ph/0609043}}].

\bibitem{Bern:1998sc}
Z.~Bern, V.~Del~Duca, and C.~R. Schmidt, {\it The infrared behavior of one-loop
  gluon amplitudes at next-to-next-to-leading order},  {\em Phys. Lett.} {\bf
  B445} (1998) 168--177, [\href{http://xxx.lanl.gov/abs/hep-ph/9810409}{{\tt
  hep-ph/9810409}}].

\bibitem{Kosower:1999xi}
D.~A. Kosower, {\it All-order collinear behavior in gauge theories},  {\em
  Nucl. Phys.} {\bf B552} (1999) 319--336,
  [\href{http://xxx.lanl.gov/abs/hep-ph/9901201}{{\tt hep-ph/9901201}}].

\bibitem{Kosower:1999rx}
D.~A. Kosower and P.~Uwer, {\it One-loop splitting amplitudes in gauge theory},
   {\em Nucl. Phys.} {\bf B563} (1999) 477--505,
  [\href{http://xxx.lanl.gov/abs/hep-ph/9903515}{{\tt hep-ph/9903515}}].

\bibitem{Bern:1999ry}
Z.~Bern, V.~Del~Duca, W.~B. Kilgore, and C.~R. Schmidt, {\it The infrared
  behavior of one-loop {QCD} amplitudes at next-to-next-to-leading order},
  {\em Phys. Rev.} {\bf D60} (1999) 116001,
  [\href{http://xxx.lanl.gov/abs/hep-ph/9903516}{{\tt hep-ph/9903516}}].

\bibitem{Catani:2000pi}
S.~Catani and M.~Grazzini, {\it The soft-gluon current at one-loop order},
  {\em Nucl. Phys.} {\bf B591} (2000) 435--454,
  [\href{http://xxx.lanl.gov/abs/hep-ph/0007142}{{\tt hep-ph/0007142}}].

\bibitem{Berends:1989zn}
F.~A. Berends and W.~T. Giele, {\it Multiple soft gluon radiation in parton
  processes},  {\em Nucl. Phys.} {\bf B313} (1989) 595.

\bibitem{Gehrmann-DeRidder:1998gf}
A.~Gehrmann-De~Ridder and E.~W.~N. Glover, {\it A complete o(alpha alpha(s))
  calculation of the photon + 1jet rate in e+ e- annihilation},  {\em Nucl.
  Phys.} {\bf B517} (1998) 269--323,
  [\href{http://xxx.lanl.gov/abs/hep-ph/9707224}{{\tt hep-ph/9707224}}].

\bibitem{Campbell:1998hg}
J.~M. Campbell and E.~W.~N. Glover, {\it Double unresolved approximations to
  multiparton scattering amplitudes},  {\em Nucl. Phys.} {\bf B527} (1998)
  264--288, [\href{http://xxx.lanl.gov/abs/hep-ph/9710255}{{\tt
  hep-ph/9710255}}].

\bibitem{Catani:1998nv}
S.~Catani and M.~Grazzini, {\it Collinear factorization and splitting functions
  for next- to-next-to-leading order {QCD} calculations},  {\em Phys. Lett.}
  {\bf B446} (1999) 143--152,
  [\href{http://xxx.lanl.gov/abs/hep-ph/9810389}{{\tt hep-ph/9810389}}].

\bibitem{Catani:1999ss}
S.~Catani and M.~Grazzini, {\it Infrared factorization of tree level qcd
  amplitudes at the next-to-next-to-leading order and beyond},  {\em Nucl.
  Phys.} {\bf B570} (2000) 287--325,
  [\href{http://xxx.lanl.gov/abs/hep-ph/9908523}{{\tt hep-ph/9908523}}].

\bibitem{DelDuca:1999ha}
V.~Del~Duca, A.~Frizzo, and F.~Maltoni, {\it Factorization of tree qcd
  amplitudes in the high-energy limit and in the collinear limit},  {\em Nucl.
  Phys.} {\bf B568} (2000) 211--262,
  [\href{http://xxx.lanl.gov/abs/hep-ph/9909464}{{\tt hep-ph/9909464}}].

\bibitem{Catani:1998bh}
S.~Catani, {\it The singular behaviour of {QCD} amplitudes at two-loop order},
  {\em Phys. Lett.} {\bf B427} (1998) 161--171,
  [\href{http://xxx.lanl.gov/abs/hep-ph/9802439}{{\tt hep-ph/9802439}}].

\bibitem{Sterman:2002qn}
G.~Sterman and M.~E. Tejeda-Yeomans, {\it Multi-loop amplitudes and
  resummation},  {\em Phys. Lett.} {\bf B552} (2003) 48--56,
  [\href{http://xxx.lanl.gov/abs/hep-ph/0210130}{{\tt hep-ph/0210130}}].

\bibitem{Mitov:2006xs}
A.~Mitov and S.~Moch, {\it The singular behavior of massive qcd amplitudes},
  {\em JHEP} {\bf 05} (2007) 001,
  [\href{http://xxx.lanl.gov/abs/hep-ph/0612149}{{\tt hep-ph/0612149}}].

\bibitem{Vermaseren:1994je}
J.~A.~M. Vermaseren, {\it Axodraw},  {\em Comput. Phys. Commun.} {\bf 83}
  (1994) 45--58.

\end{thebibliography}\endgroup
\bibliographystyle{/home/stefanw/latex-style/JHEP.bst}

\end{document}